\documentstyle[epsf,rotate]{elsart} 
\begin{document}

\begin{frontmatter}

\title{Thermostatistics of extensive and non--extensive systems using
generalized entropies}
\author{R. Salazar\thanksref{email1}} 
\author{and R. Toral\thanksref{email2}}  
\thanks[email1]{e-mail: rafael@imedea.uib.es}
\thanks[email2]{e-mail: raul@imedea.uib.es} 
\address{Instituto Mediterr\'aneo de Estudios Avanzados  
(IMEDEA\thanksref{web1}, UIB--CSIC) and Departament de 
F\'{\i}sica, Universitat de les Illes Balears,  
07071 Palma de Mallorca, Spain} 
\thanks[web1]{URL: http://www.imedea.uib.es}

\begin{abstract} 
We describe in detail two numerical simulation methods valid to study
systems whose thermostatistics is described by generalized entropies, such
as Tsallis. The methods are useful for applications to non-trivial
interacting systems with a large number of degrees of freedom, and both
short--range and long--range interactions. The first method is quite
general and it is based on the numerical evaluation of the density of
states with a given energy. The second method is more specific for Tsallis
thermostatistics and it is based on a standard Monte Carlo Metropolis
algorithm along with a numerical integration procedure. We show here that
both methods are robust and efficient. We present results of the
application of the methods to the one--dimensional Ising model both in a
short--range case and in a long--range (non--extensive) case. We show that
the thermodynamic potentials for different values of the system size $N$
and different values of the non--extensivity parameter $q$ can be described
by scaling relations which are an extension of the ones holding for the
Boltzmann--Gibbs statistics ($q=1$). Finally, we discuss the differences in
using standard or non--standard mean value definitions in the Tsallis
thermostatistics formalism and present a microcanonical ensemble
calculation approach of the averages.

\end{abstract}


\end{frontmatter}

\section{Introduction} \label{s1}
Non--extensive systems are those for which the thermodynamic potentials do
not scale linearly with the system size. As a way of example, in some
electric or magnetic systems with very long--range interactions the ground
state energy per particle increases with the number of particles. In the
absence of other effects, such as screening, these system are ``genuinely"
non--extensive. If we apply to them the standard Boltzmann--Gibbs formalism
of the Statistical Mechanics, we find that the internal energy, Helmholtz
free energy and other thermodynamic potentials are non--extensive as well.
This standard formalism can be implemented by using the definition of the
entropy $S$ in terms of the probabilities $p_i$ of the $i=1,\dots,W$
possible microscopic configurations\footnote{We use throughout the paper
dimensionless units where the Boltzmann constant, $k_B$ is equal to $1$}:

\begin{equation}\label{e0} 
S= -\sum_{i} p_i \ln p_i
\end{equation} 

The actual calculation of the entropy assumes a set of probabilities
$\{p_i\}$. These are computed by finding the maximum of the above
expression when some extra conditions defining an ensemble (fixed number of
particles and mean energy, for example) are imposed. 

Even for systems in which the energy levels do scale with the system size,
it is possible, by using generalized definitions of the entropy, to obtain
non--extensive thermodynamic potentials. One of the most successful
generalizations is that of Tsallis which in 1988 proposed \cite{tsa88} the
following alternative expression for the entropy:

\begin{equation}\label{e1} 
S_q=\frac {1- \sum_{i} p_i^q}{q-1} 
\end{equation} 

where $q$ is an entropic index that characterizes the {\it degree of
non--extensivity}. It is possible to show that the entropy of the composed
system A+B satisfies the relation:

\begin{equation}\label{e2}
S_q(A+B)=S_q(A)+S_q(B)+(1-q)S_q(A)S_q(B) 
\end{equation}

when A and B are independent systems in the sense that $p_{ij}(A+B)= p_i(A)
p_j(B)$. We see that for $q \ne 1$ there is no additivity in the entropy,
which also implies non--extensivity. The Boltzmann--Gibbs entropy, Eq.
(\ref{e0}), and extensivity  are recovered in the limit $q \to 1$. Since
the probabilities $\{p_i\}$ satisfy $p_i^q>p_i$ for $q<1$ and $p_i^q<p_i$
for $q>1$, the superextensive, $q<1$, and the subextensive, $q>1$, regimes
will privilege the {\it rare} and {\it frequent} events respectively.

In the last years there have been many studies in which Tsallis
non--extensive statistics has been applied to different situations (See
\cite{tsa95,tsa99} for a review). In some cases, the systems considered are
genuinely non--extensive (in the sense defined above) while in others the
non--extensivity arises as a result of the application of the new
statistics. In fact, and due to the intrinsic non--extensivity of the
Tsallis statistics, it has been argued that its natural range of
applicability should include systems with long--range interactions or
long--range microscopic memory processes, as well as dynamical systems in
which the space--time geometry has a multifractal--like structure, because
those systems are in general genuinely non--extensive. Although most of the
literature (including this paper) basically derives equilibrium properties
starting from the generalized definition of the entropy, it has been
conjectured recently \cite{tsa99}, however, that the Tsallis entropy could
be relevant instead in the study of non--equilibrium processes.

Due to the difficulty of deriving exact results, it is natural to use
numerical methods to obtain the properties of a  system with many degrees
of freedom when studied under the rules of the new statistics. This is
necessary in order to extract results that could be checked against
experiments. However, these studies have been hampered by the failure of
the typical Monte Carlo methods to adequately generate representative
equilibrium configurations distributed according to Tsallis statistics. It
is the purpose of this paper to explain in detail new methods that can be
used to study the equilibrium properties of a many--particle system when it
is considered under generalized statistics. Although our methods are quite
general, we will illustrate their use by considering a prototypical
genuinely non--extensive system: the Ising ferromagnet model with
long--range interactions. We will also consider the short--range Ising
model in order to test the simulation methods and to compare the results
obtained from the use of Tsallis statistics in extensive and non--extensive
systems. 

In the remaining of the section, we will outline briefly which are the
basic difficulties one encounters when trying to generalize the standard
Monte Carlo methods (such as the Metropolis algorithm) to the study of
Tsallis statistics. The main problem is that the probabilities $\{p_i\}$
can not be given an explicit expression, as we will see in the following
discussion. Let us consider the canonical ensemble. The probabilities
$\{p_i\}$ in this ensemble are found by solving the maximization problem of
the entropy  $S_q$ as given by Eq.(\ref{e1}) subject to the constrains of
(i) positivity: $p_i\ge 0$, (ii) normalization: $\sum_{i} p_i=1$ and (iii)
a fixed mean value for the internal energy:  $\langle {\cal H} \rangle =U$,
where $\cal H$ is the Hamiltonian of the system and the mean value of any
function $O$ of the microscopic configurations is computed according to the
general rule:

\begin{equation}
\label{eq4}
\langle O \rangle =\sum_{i} O_i u(p_i),
\end{equation} 

$O_i$ is the value of $O$ at the configuration whose probability is $p_i$
and we have introduced a function $u(p_i)$ that allows the definition of
generalized mean values. The standard mean values are recovered by taking
$u(p_i)=p_i$.  Although, initially, the choices $u(p_i)=p_i$ ({\sl first
option}), and $u(p_i)=p_i^q$ ({\sl second option}) were considered, later,
it was shown that a better choice, in the sense that it preserves the
Legendre structure of the resulting thermodynamics formalism, is to
consider $u(p_i)=p_i^q/\sum_jp_j^q$ ({\sl third option}) \cite{tsa98}. We
will use the following notation of the averages in this third option:

\begin{equation}\label{e3} 
\langle O\rangle_q = \sum_{i} O_i P_i~;~~~~~~~~~~
P_i = \frac {p_i^q}{\sum_{j} p_j^q}. 
\end{equation} 
 
$\{P_i\}$ are known as the ``escorts" probabilities \cite{bec93}. It is
possible to recover the configuration probabilities $p_i$ from the escorts
probabilities using:

\begin{equation}
p_i=\frac{P_i^{1/q}}{\sum_{j} P_j^{1/q}}.
\end{equation}

The entropy, in terms of the $\{P_i\}$'s is given by:

\begin{equation}\label{e5} 
S_q=\frac {1- (\sum_{i} P_i^{1/q})^{-q}}{q-1} 
\end{equation} 

Concerning the different definitions for the averages, it should be said
that it has been shown recently \cite{men97,pla97b,gue96} that the standard
mean values of the first option, $u(p_i)=p_i$, can be also made compatible
with the Legendre structure of the  thermodynamics and the resulting
formalism also represents a thermodynamically stable description. In this
paper, we follow mainly the formulation in terms of the mean values defined
by Eq.(\ref{e3}), although in a later section we will show that the results
obtained using the standard mean values can be mapped onto the ones
obtained using Eq. (\ref{e3}).

The maximization problem for the unknown escort probabilities $P_i$ in the
canonical ensemble with a given internal energy $U_q$ is:

\begin{eqnarray} 
\frac{\delta~~}{\delta P_i} \bigl[ S_q-\beta \sum_i\varepsilon_i P_i -\alpha  
\sum_i P_i \bigr] & = & 0\label{e2a}\\
P_i & \ge  & 0\label{e2b}\\
\sum_{i} P_i& = & 1 \label{e2c}\\
\sum_{i} \varepsilon_i P_i & = & U_q\label{e2d}
\end{eqnarray}

where $\alpha$, $\beta$ are Lagrange multipliers. $\{\varepsilon_i\}$ are
the energy levels of the system under consideration whose ground state
energy will be denoted by $E_0$.   Solving the problem
Eq.(\ref{e2a}-\ref{e2d}) one obtains the probabilities for the canonical
ensemble as \cite{tsa98}: 

\begin{equation}\label{e7} 
P_i = \left \{ \begin{array}{lr}
0, & 1-\frac {(1-q) \beta (\varepsilon_i-U_q)}{(\sum_{j} P_j^{1/q})^q} <  0\\
\frac {[1-(1-q) \beta (\varepsilon_i-U_q)/(\sum_{j}
P_j^{1/q})^q]^{\frac q {1-q}}}  {\sum_{k} [1-(1-q) \beta
(\varepsilon_k-U_q)/(\sum_{j} P_j^{1/q})^q]^{\frac q {1-q}}}, &{\rm otherwise}
\end{array} \right.
\end{equation} 

The probabilities defined in this way are real and non--negative. The
condition giving the possible values of  $\beta$ and $\varepsilon_i$  for
which  $P_i \ne 0$ in Eq. (\ref{e7}) is called the {\it cut--off
condition}.  One can show that the probabilities Eq. (\ref{e7}) are
invariant under a change in the energy levels $\varepsilon_i \to
\varepsilon_i +\Delta\varepsilon$  (and the same change in the internal
energy $U_q\to U_q+\Delta\varepsilon$) for arbitrary $\Delta \varepsilon$.
By introducing the temperature $T=1/\beta$, it is possible to show also the
validity of the relation \cite{pla97b,tsa98}

\begin{equation}\label{e8} 
1/T=\partial S_q /\partial U_q 
\end{equation} 

which reflects the Legendre structure of the thermodynamics obtained.

Notice that Eq.(\ref{e7}) does not give yet the actual values of the
probabilities since the $\{P_i\}$'s appear in a non--trivial way in both
sides of the equation (either explicitly or in the cut-off condition). This
is different from the solution obtained in the usual Boltzmann--Gibbs
canonical ensemble (recovered in the limit $q\to 1$) in which the solution
adopts the explicit form:

\begin{equation}
\label{eq12}
P_i=\frac{{\rm e}^{-\beta \varepsilon_i}}
{\sum_{j}{\rm e}^{-\beta \varepsilon_j}}
\end{equation}

(although, of course, it is very difficult to compute the denominator of
this expression, the partition function, for an interacting system). An
iterative method to solve Eq.(\ref{e7}) has been used in \cite{tsa98}. In
this method, an initial guess for the probabilities is fed in the right
hand side of (\ref{e7}) and this equation is used recursively until
convergence is achieved. We will see, however, that for many--particle
systems, it might very difficult to achieve convergence in some cases. 

A convenient way of writing Eq. (\ref{e7}) is by using an auxiliary
parameter $\beta'$ defined as \cite{tsa98}:

\begin{equation}\label{e10} 
\beta'= \frac {\beta}{(1-q)\beta \sum_{j} \varepsilon_j P_j +
(\sum_{j} P_j^{1/q})^{-q}} 
\end{equation}

Defining  $T'\equiv 1/\beta'$, and using Eqs. (\ref{e5}) and (\ref{e2d})
one can rewrite this equation as

\begin{equation}\label{e11} 
T= \frac {T' - (1-q) U_q}{1 + (1-q) S_q} 
\end{equation}

In terms of $\beta'$ the solution adopts a form similar to that of the
standard canonical ensemble:

\begin{equation}\label{e9}
P_i = \left \{ \begin{array}{lr}
0, & 1- (1-q) \beta' \varepsilon_i <  0\\
\frac {[1-(1-q) \beta' \varepsilon_i]^{\frac q {1-q}}}  
{\sum_{j} [1-(1-q) \beta' \varepsilon_j]^{\frac q {1-q}}}, & {\rm otherwise}
\end{array} \right.
\end{equation} 

One can then adopt the following practical procedure \cite{tsa98}: choose
a value for the parameter $T'$ and compute the probabilities $P_i$ as a
function of $T'$ using Eq.(\ref{e9}). Compute the internal energy $U_q$,
the entropy $S_q$ and the temperature $T$, always as a function of $T'$,
using Eqs. (\ref{e2d}), (\ref{e5}) and (\ref{e11}), respectively. Finally,
vary $T'$ in order to make the parametric plots $U_q(T)$ and $S_q(T)$.
Other thermodynamic potentials follow the usual definition. For instance,
the Helmholtz free energy is $F_q=U_q-TS_q$. 

It is important to realize that although the probabilities $P_i$, when
considered as a function of $T$, do not depend on an arbitrary shift
$\Delta\varepsilon$ of the energy levels or, in other words, do not depend
on the zero of energy, $E_0$, they do depend on $E_0$ when considered as a
function of $T'$. This means that the averages as a function of $T'$ can
not be physically relevant {\sl because they depend on the zero of energy}.
This is why $T'$ has to be interpreted only as an auxiliary parameter, not
as an actual temperature. Of course, the relation $T'\to T$ depends also on
the zero of energy in such a way that both dependences cancel and the
averages as a function of $T$ are independent of a shift in the energy
levels. An interesting question is to determine the range of values for the
parameter $T'$ that should be used in order to obtain the usual range $0
\le T < \infty$ for the temperature $T$. Using that, according to the
definition (\ref{e1}), it is $1+(1-q)S_q > 0$ we obtain from Eq.(\ref{e11})
that $T'$ should vary in the range $[(1-q)E_0,\infty)$, where we have used
that the energy at zero temperature is the ground state energy
$U_q(T=0)=E_0$. Therefore, it is important to use the right range of values
for $T'$ in order to reach all the possible values for $T$. In particular,
$T'$ might need to take negative values either for $q<1$ or for $q>1$
unless one adopts $E_0=0$ as we will throughout this paper. To our
understanding, it is not clear in the literature the fact that the averages
as a function of the parameter $T'$ depend on the zero of energy and that
it might be necessary to consider negative values for $T'$ in order to span
the whole range of values for $T$.

As stated before, the main problem to  perform Tsallis thermostatistics
simulations at a given temperature $T$ is that there is not an {\sl
explicit} expression for the probabilities $P_i$, c.f. see Eq. (\ref{e7}).
The practical procedure outlined above (compute $P_i$ as a function of $T'$
and then compute $U_q$, $S_q$ and $T$ as a function of $T'$ in order to
make parametric plots by varying $T'$) is not straightforward to implement
numerically since it is very difficult to use Eq.(\ref{e5}) to compute the
entropy. This is because the usual Monte Carlo methods, i.e. the Metropolis
algorithm, require only the probabilities $P_i$ up to a normalization
factor where Eq.(\ref{e5}) requires the absolute, normalized values of
$P_i$. It is the object of this paper to explain in detail some numerical
methods of the Monte Carlo type that can be used to perform the necessary
averages for generalized statistics, including Tsallis.

There have been previous attempts to perform numerical simulations of
Tsallis statistics using Monte Carlo methods. An earlier work in this
direction is that of T. Penna {\it et. al.} \cite{pen95} who extended the
Metropolis acceptance procedure to include a dependence in the $q$
parameter. However, this method does not satisfy the detailed balance
condition which is a key ingredient of Monte Carlo methods. Another
interesting approach is that of I. Andricioaei {\it et. al.} \cite{ioa96},
who performed a Metropolis Monte Carlo algorithm which does satisfy the
detailed balance condition for the probability $P_i$ as a function of the
parameter $T'$ but, since  they do not make the temperature transformation
$T' \to T$, they are unable to determine the actual temperature $T$ of the
simulation. All these works considered the second version, $u(p_i)=p_i^q$,
for the definition of averages, which, as discussed before, has proven
afterwards not to be the optimal election \cite{tsa98}. We have used also a
similar sampling in the context of Simulated Annealing \cite{sal98d,sal97}.
A recent approach proposed by A. R. Lima {\it et. al.} \cite{lim99}  uses
the {\it broad  histogram Monte Carlo method}, which determines the number
of microstates using a balance equation between near neighbor energy
levels. They are able to apply this method to the Ising Model with
short--range interactions. This is a valid Monte Carlo simulation with full
control of the temperature $T$ but its applicability is somewhat
restricted. As we will show, the Ising model with long--range interactions
can not be treated straightforwardly with this method because the spin flip
dynamics does not produce transitions between near energy levels and,
consequently, the broad histogram Monte Carlo method, in its present form,
can not be used to study long--range interacting systems. We are able to
overcome these difficulties by extending a method which had been developed
some time ago \cite{bha87,bha87b,bha87c,bha87d}  to compute directly the
number of states with a given energy and which does not depend on the
definition of the entropy. In fact, our method can be used to study any
statistics and applications will be shown both for the Boltzmann--Gibbs and
the Tsallis statistics. We develop yet a second method which is devised
specifically for the Tsallis statistics and that has the advantage that it
uses the familiar Metropolis algorithm plus a numerical integration.

This paper is organized in the following form: in section \ref{s2} we use a
simple and limited enumeration procedure valid only for small system 
sizes. However, this method is exact and can be used to check against some
of the approximated methods we will introduce later. In this section we
also introduce the short--range Ising model (SRIM) and the long--range
Ising Model (LRIM). These two models will be employed in this paper in
order to test the numerical methods described here and to compare the
behavior of the non--extensive Tsallis thermostatistics in genuinely
extensive and non--extensive systems. In section \ref{s4} we explain in
some detail the Histogram by Overlapping Windows (HOW) method. We show that
this method has a wide range of applicability since it can be used both for
short--range and long--range systems as well as for any kind of statistics.
Section \ref{s3} presents a Metropolis Monte Carlo type method specially
devised to study systems in the Tsallis statistics. In section \ref{s5} we
present some results concerning the validity of some  scaling relations for
the SRIM and LRIM in the Tsallis thermostatistics context. These scaling
relations are an extension of the ones holding for the Boltzmann--Gibbs
statistics . In section \ref{s6} we present results for these models using
standard mean values instead of those defined by  Eq. (\ref{e3}). In
section \ref{s7} we discuss the results of using Tsallis statistics in the
microcanonical ensemble. Finally, in section \ref{s8} we summarize the main
conclusions of this work.

\section{Exact enumeration}\label{s2}

The problem to compute the probabilities $\{P_i\}$ using Eq.(\ref{e9}) is
that the number of terms in the sum of the denominator of this equation,
the number of microstate configurations $W$, is extremely large (typically
scales exponentially with the system size). However, for small systems, it
might be possible to enumerate completely the microstates and, therefore,
to compute magnitudes of interest such as the internal energy $U_q(T)$, the
entropy $S_q(T)$, the free energy $F_q(T)$, etc. We follow this approach in
this section. Although the system sizes one can usually study with this
method are very far from reaching a situation in which scaling laws with
system size apply, we use the results as a bench-test in order to compare
with  other approximate methods that will be introduced in the following
sections.

We will consider Ising type models with Hamiltonian:

\begin{equation}\label{e14} 
{\cal H} = \sum_{(i,j)} J_{ij} ({1-s_i s_j}) 
\end{equation} 

where each of the $N$ spin variables $s_i$, $i=1,\dots,N$ can take the
values $\pm 1$. The sum $\sum_{(i,j)}$ runs over all distinct pairs of
sites on a $d$-dimensional regular lattice of linear size $L=N^{1/d}$ with
periodic boundary conditions and lattice constant equal to $1$. $J_{ij}$ is
the coupling parameter between spins $i$ and $j$. Note that for
ferromagnetic couplings, $J_{ij}\ge 0$, the ground state is double
degenerate and its energy is $E_0=0$. The usual, nearest--neighbors, or
short--range Ising model (SRIM), is obtained taking $J_{ij}=1$, if
$r_{ij}=1$ and $J_{ij}=0$, if $r_{ij}>1$. The long--range Ising model
(LRIM) is defined by using

\begin{equation}
J_{ij}=1/r_{ij}^{\alpha},
\end{equation}

where $r_{ij}$ is the  distance between the spins $i$ and $j$, and the
parameter $\alpha$ sets the interaction range.  The SRIM is formally
recovered by taking the limit $\alpha \to \infty$. Depending on the value
of $\alpha$ and the space dimension $d$, the LRIM has two regimes: the
extensive regime, $\alpha>d$, and the non--extensive regime, $\alpha\le d$.
This can be seen by roughly estimating the mean energy per spin in an
infinite system as   $\int_1^{\infty} dr r^{d-1}r^{-\alpha}$. We obtain a
convergent integral if $\alpha>d$ (extensive behavior), and for $\alpha \le
d$ the integral diverges (non--extensive). More precisely, a convenient
scale for the mean energy per spin in a finite system of size $N$ is given
by \cite{jun95}:

\begin{equation}\label{e28}
\tilde N = 1+d \int_1^{L} dr r^{d-1}r^{-\alpha}= 
\frac{N^{1-\alpha/d}-\alpha/d} {1-\alpha/d} 
\end{equation}

The definition of $\tilde N$ is such that the limit $\alpha \to d $ is well
defined. Again, we see that for $\alpha>d$ the internal energy per spin
scales as a constant in the limit of large $N$, but for $\alpha \le ad$, it
grows with the system size. The system is, in this latter case, genuinely
non--extensive. The SRIM limit, $\alpha \to \infty$, gives the expected
result $\tilde N = 1$.

\begin{figure}[!ht] 
\centerline {\epsfxsize=10cm \epsfbox{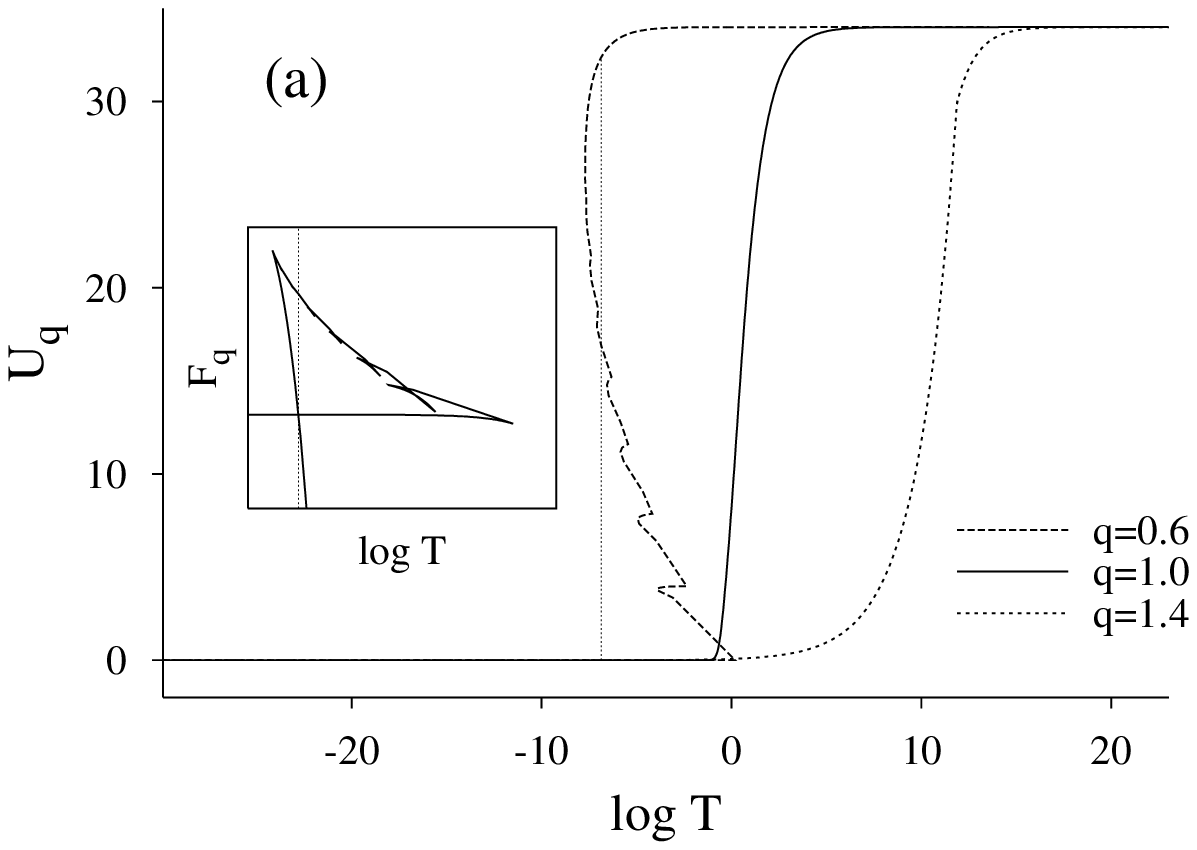}}
\centerline {\epsfxsize=10cm \epsfbox{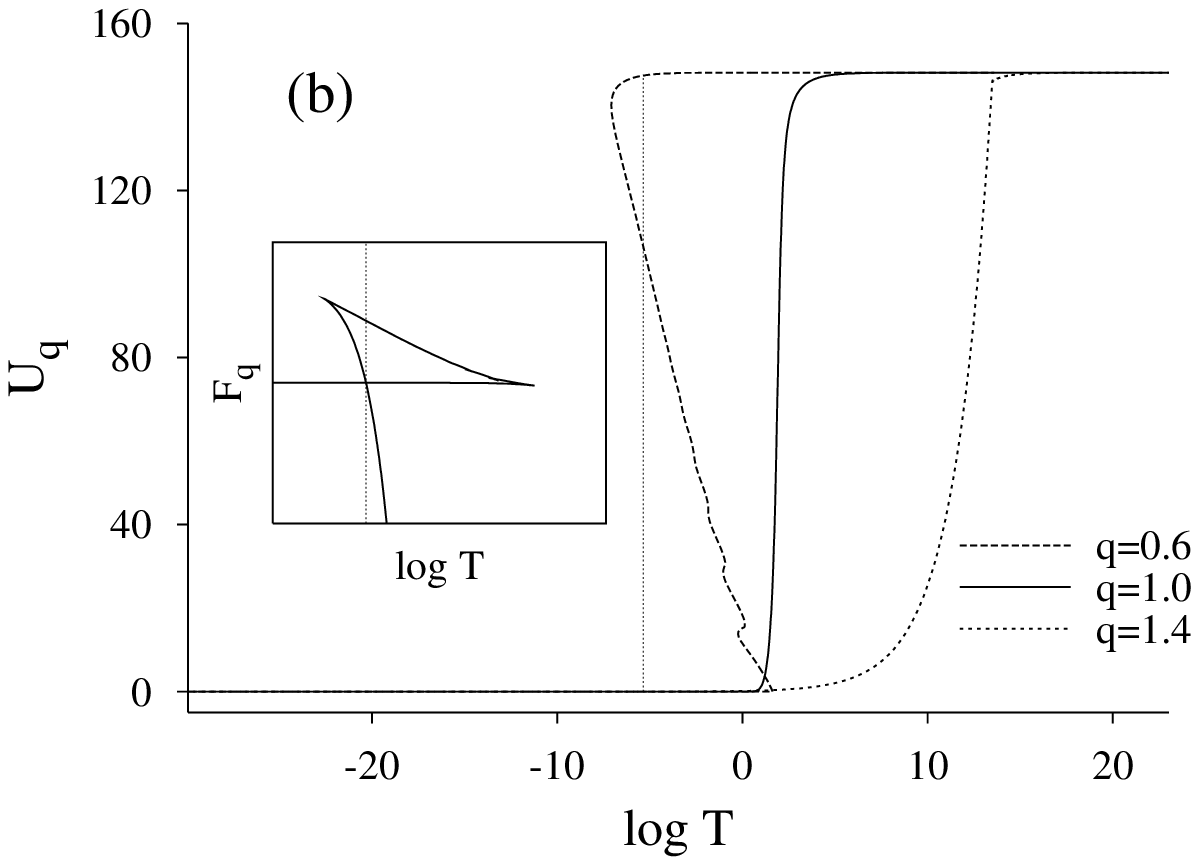}}
\caption{Internal energy $U_q$ as a function of the temperature $T$, from
the exact evaluation of Eqs. (\ref{e2d},\ref{e7}) for one--dimensional
Ising models with $N=34$ spins. Plot (a) is for the SRIM and plot (b) for
the LRIM in a genuinely long-range case, $\alpha=0.8$. The insert shows the
free energy $F_q$ for $q=0.6$. In this case, a part of the energy curve is
replaced by a vertical straight line (shown by dots in the plot) such that
the Maxwell criterion of equal free energies is satisfied. 
\label{f1}}
\end{figure}

The number of configurations in the Ising model is $W=2^N$. We have made a
complete enumeration of the $i=1,\dots,W$ configurations and their energies
$\varepsilon_i$ for a linear chain, $d=1$, of sizes up to $N=34$. We have
used these results to compute the probabilities $P_i$ and then the
thermodynamic magnitudes of interest using Tsallis statistics with the
third option for the averages. In Fig. \ref{f1}, we plot the exact internal
energy $U_q$ as a function of the temperature $T$,  for several values of
the parameter $q$ for the LRIM in a genuinely non--extensive situation,
$\alpha=0.8$, Fig. \ref{f1}.b, and for the SRIM, Fig. \ref{f1}.a. In the
$q<1$ case we observe that there is a range of temperatures for which the
internal energy is not a single--valued function of the temperature and,
for a given value of $T$, there are several possible values for $U_q$. This
ambiguity is resolved by using a Maxwell--like construction
\cite{lim98,tsa98} that replaces the loop in the energy curves by a
vertical straight line connecting the two points with the same free energy
$F_q(T)$.

\begin{figure}[!ht] 
\centerline {\epsfxsize=10cm \epsfbox{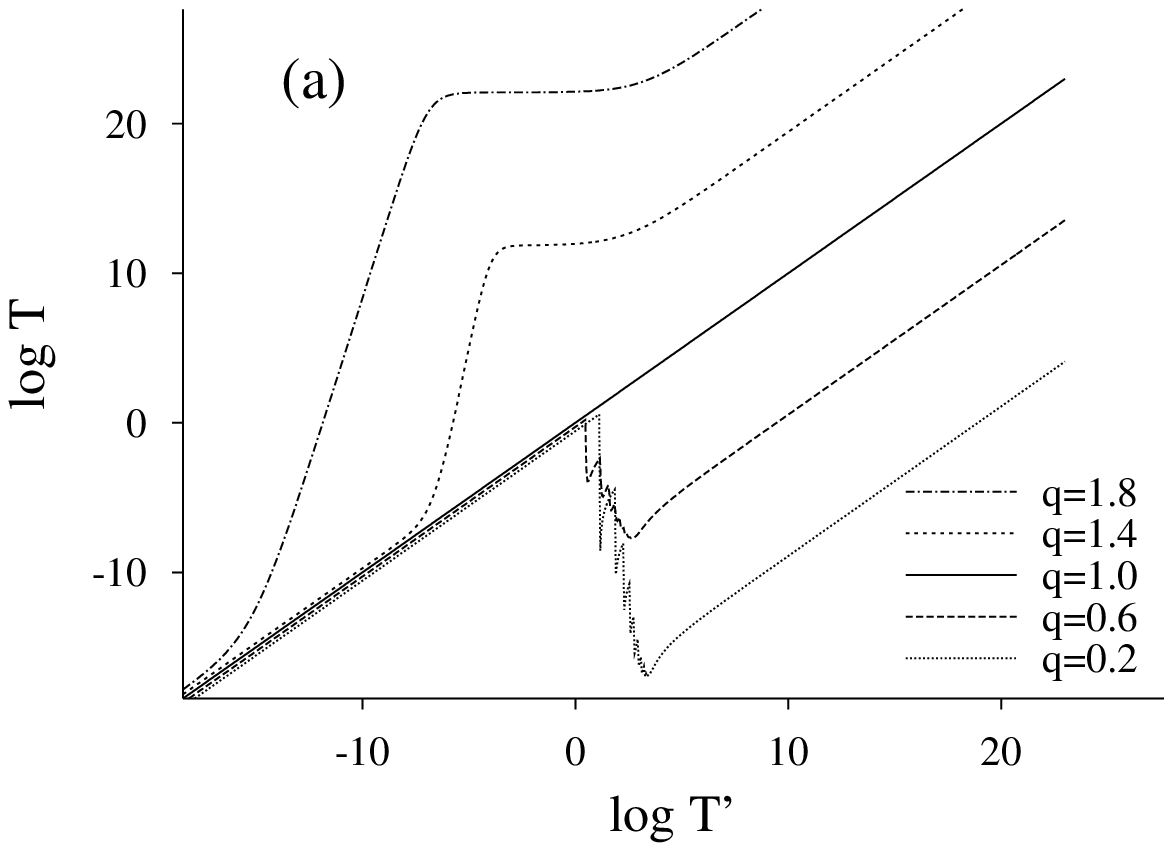}}
\centerline {\epsfxsize=10cm \epsfbox{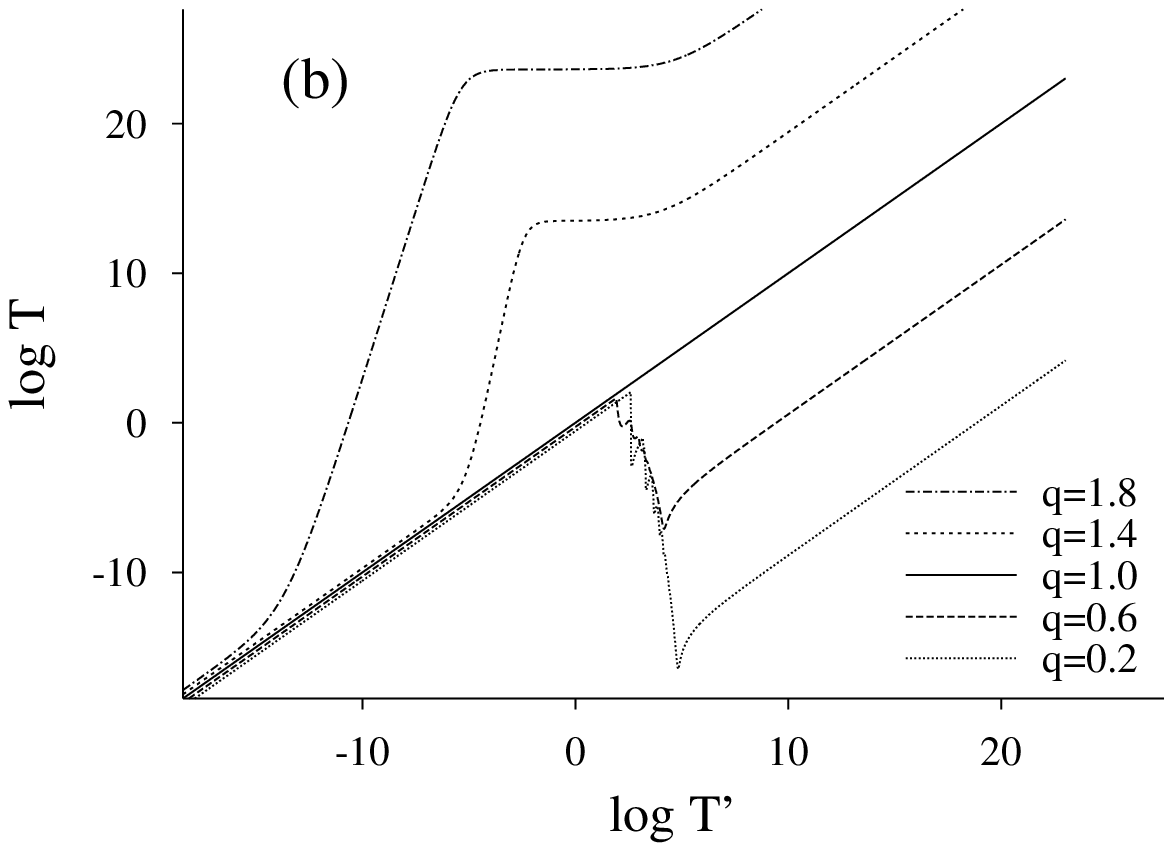}}
\caption{The $T' \to T$ transformation for the one--dimensional Ising
models as obtained from the exact evaluation of Eq. (\ref{e11}) for a
system size $N=34$ and different values of $q$. Plot (a) is for the SRIM,
while plot (b) is for the LRIM for $\alpha=0.8$. Notice that, in both
cases, there is a temperature range for which the curves are almost
horizontal. This makes it very difficult to use iterative methods for the
determination of the probabilities using directly Eq. (\ref{e7}). 
\label{f2}}
\end{figure}

We stress that the loops in the energy curves appear as a result of the
$T'\to T$ transformation and, therefore, will not be observed when plotting
the energy as a function of $T'$. Typical $T' \to T$ transformations are
shown in Fig. \ref{f2}. We observe in this figure that for $q<1$ a same
value of $T'$ can correspond, in some cases, to three or more values of $T$
giving rise to the observed multivalued behavior in the energy. Figure
\ref{f2} helps us to understand the failure of the iterative method that
has been proposed \cite{tsa98} to solve the set of equations (\ref{e7}).
Although each value of $T'$ defines a unique set $\{P_i\}$. We see Fig.
\ref{f2} that for $q>1$ there are some intervals of $T'$ where the
transformation $T' \to T$ is almost horizontal. Therefore, one value of $T$
corresponds nearly to a complete interval for $T'$ and hence, there are
many possible solutions for $\{P_i\}$ very close to the real one. This is
the main reason for the failure of iterative methods for $q>1$. The
situation worsens for increasing system size $N$.

Whatever illuminating the method of exact enumeration is, its validity is
limited to very small values for $N$. To our knowledge, the largest value
ever considered in an exact enumeration scheme for a short--range Ising
model is $N=4^3=64$ \cite{pea82}.  Simulations at larger $N$ sizes require
other methods, as the ones presented in the next two sections.

\section{The Energy Histogram Method using Overlapping Windows}\label{s4}

Although the number of possible microscopic configurations is in general
very large, the range of possible energy values usually takes a much
smaller value. For instance, for the one--dimensional SRIM introduced in
the previous section, with $N$ spins we have $W=2^N$, but the number of
possible energy values is $N/2+1$. Let $M$, in general, be the number of
possible energy levels. We denote by $\Omega(E_k)$ the number of
microscopic states sharing the same energy $E_k$, $k=0,\dots,M-1$.
Obviously $\sum_{k}\Omega(E_k)=W$. We  rewrite all the sums in Eqs.
(\ref{e9},\ref{e5},\ref{e3}) as:

\begin{eqnarray} 
P(E_k) &=&  \left \{ \begin{array}{lr}
0, & 1- (1-q) \beta' E_k <  0\\
\frac {[1-(1-q) \beta' E_k]^{\frac q {1-q}}}  
{\sum_{n} \Omega(E_n) [1-(1-q) \beta' E_n]^{\frac q {1-q}}}, & {\rm otherwise}
\end{array} \right. \label{e22a}\\  
S_q &=& \frac {1- (\sum_k \Omega(E_k) P(E_k)^{1/q})^{-q}}{q-1} \label{e22b}\\  
\langle O\rangle_q &=& \sum_k \Omega(E_k) O(E_k)
P(E_k) \label{e22c}
\end{eqnarray} 

where the sums run over the $M$ energy levels.

Notice that, once the $\Omega(E_k)$'s have been computed, {\sl any
statistics can be performed upon the system}. Whether we use Tsallis,
Boltzmann--Gibbs or any other generalized statistics is simply a trivial
change in the computational scheme. Moreover, it is also trivial to compute
the averages for any value of the parameters, say $T$ or $q$. Therefore,
although the calculation of the $\Omega_k$'s might be time consuming, the
pay--off is tremendous\footnote{All the simulations in this paper have been
performed using a Pentium-III processor at 550MHz}.

In general, the $\Omega(E_k)$ are very difficult to obtain exactly. An
important exception that will be used throughout this paper is the SRIM in
1-d, for which the energy levels are given by $E_k=4k$ for $k=0,\dots,N/2$
and whose degeneracy is:

\begin{equation}
\label{eq:24}
\Omega(E_k)=2{N\choose{E_k/2}}
\end{equation}

In the cases in which $\Omega(E_k)$ is not known we need approximate
numerical methods. The most naive way to find the $\Omega(E_k)$'s is to
generate different system configurations randomly and count how many times
a configuration with energy $E_k$ appears. However, this approach fails
because the complete set of $\Omega(E_k)$ values span too many orders of
magnitude. In general, two energy levels $E_k$ and $E_n$ could differ as
much as $\Omega_k/\Omega_n  \sim \exp{N}$. This means that the range of
variation of $\Omega(E_k)$ over the $M$ different energy levels is very
large and it is not possible to generate in a single run a histogram that
covers all the energy levels, unless one generates a number of
configurations of the order of the total number available to the system,
$W$.

The Histogram by Overlapping Windows method (HOW) used here \cite{bha87}
avoids this problem by generating system configurations within a restricted
energy interval and estimating the relative weights of these energy levels
from the number of times they appear in each interval. By generating enough
intervals spanning the whole energy range, one is able to obtain good
quality estimators of the numbers $\Omega(E_k)$. An earlier account of the
method has been given in \cite{sal00} and we explain now in some detail how
the method works.

Let us consider first the SRIM in arbitrary dimension. In this case, the
possible energy values are $E_k=4k$ for $k=0,\dots,dN/2$. Following the
original work \cite{bha87}, we consider the intervals (windows)
$\{E_0,E_1,E_2,E_3\}$, $\{E_3,E_4,E_5,E_6\}$, $\{E_6,E_7,E_8,E_9\}$, etc.
Each window consists of $4$ consecutive energy levels and the last energy
value of one window is the first of the next one. The next step is to take
one of the intervals and to generate configurations whose energy belongs to
it. This is achieved, after preparing the system initially with one of the
energies of the interval, by flipping spins chosen at random. A spin flip
is accepted only if it leaves the system in one of the energy levels of the
interval and it is rejected otherwise. The ratio of the number of generated
states with energy $E_k$ to the number of generated states with energy
$E_n$ is an unbiased estimator of $\Omega(E_k)/\Omega(E_n)$, for $E_k$ and
$E_n$ within the energy window. The quality of the estimator increases with
the number of generated configurations. From the overlap between windows
one can compute $\Omega(E_k)$ for the whole range of energies. The number
of energy values in each window ($4$ in the previous example) is not
important as far as it is not too large (such that the ratios
$\Omega(E_k)/\Omega(E_n)$ are not extremely small or large) and it is not
too small either. If the window is very small, most spin flips will yield
an energy outside the range of allowed values and the number of accepted,
i.e. independent, configurations will be very small. Moreover, the final
algorithm must be ergodic: any energy value in a window should be obtained
from any other value in the same window after a sufficient number of spin
flips.

\begin{figure}[!ht]  
\centerline {\epsfxsize=10cm \epsfbox{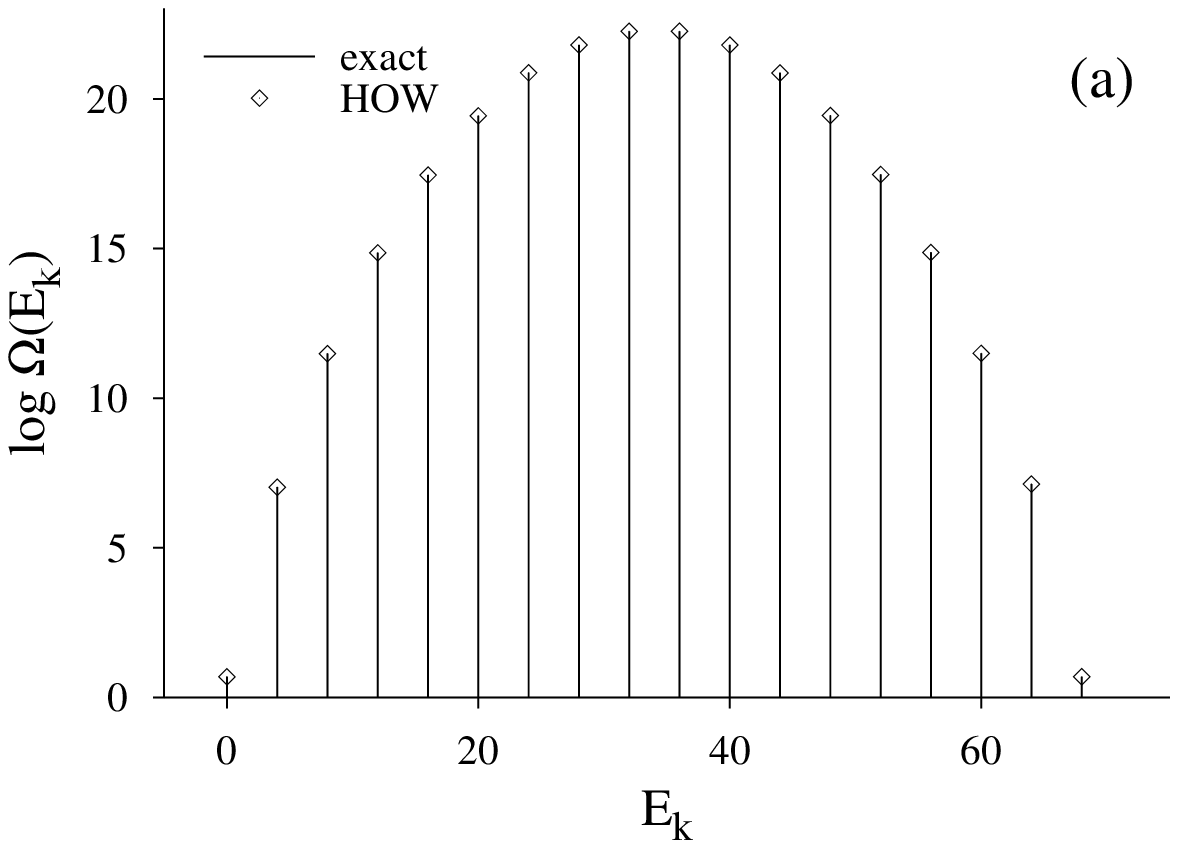}}
\centerline {\epsfxsize=10cm \epsfbox{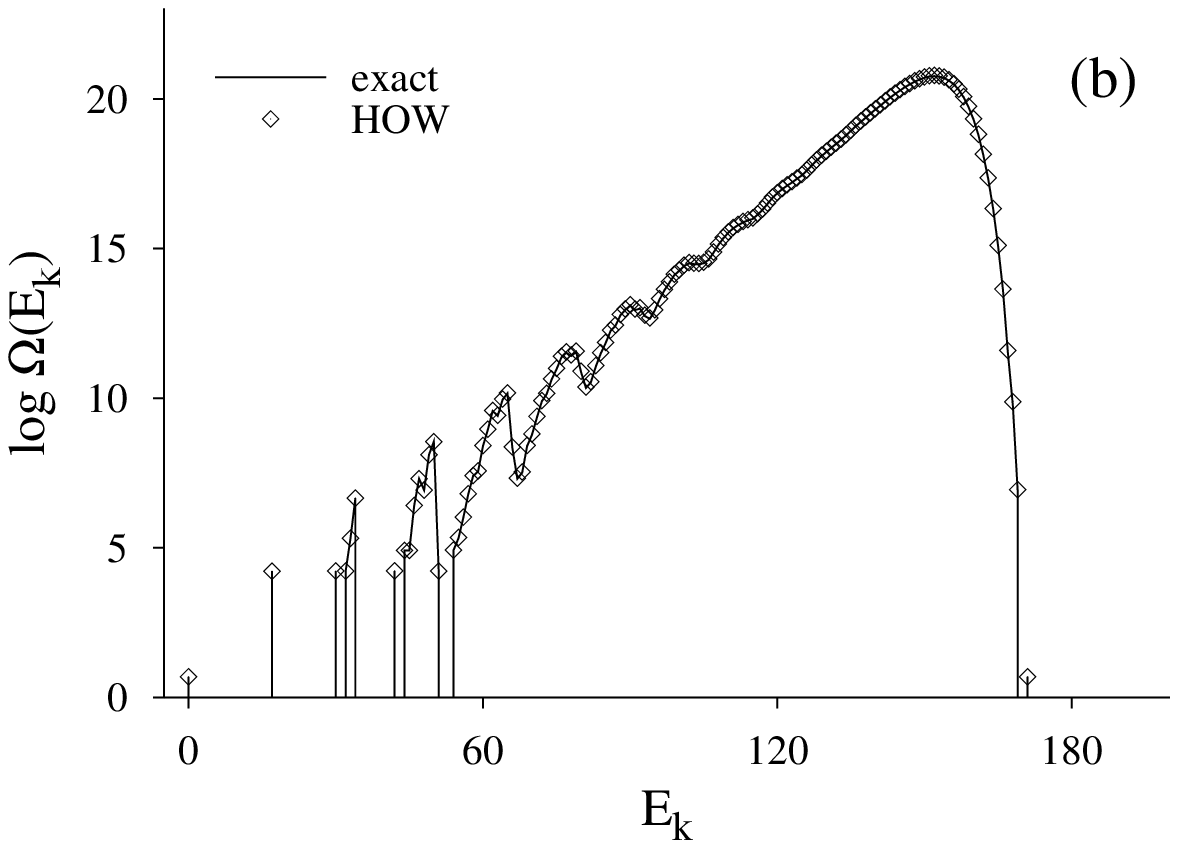}}
\caption{The number of states $\Omega(E_k)$ for $N=34$ calculated exactly
(lines), and using the HOW method (symbols). (a) SRIM, (b) LRIM.
\label{f3}}
\end{figure}

\begin{figure}[!ht]  
\centerline {\epsfxsize=10cm \epsfbox{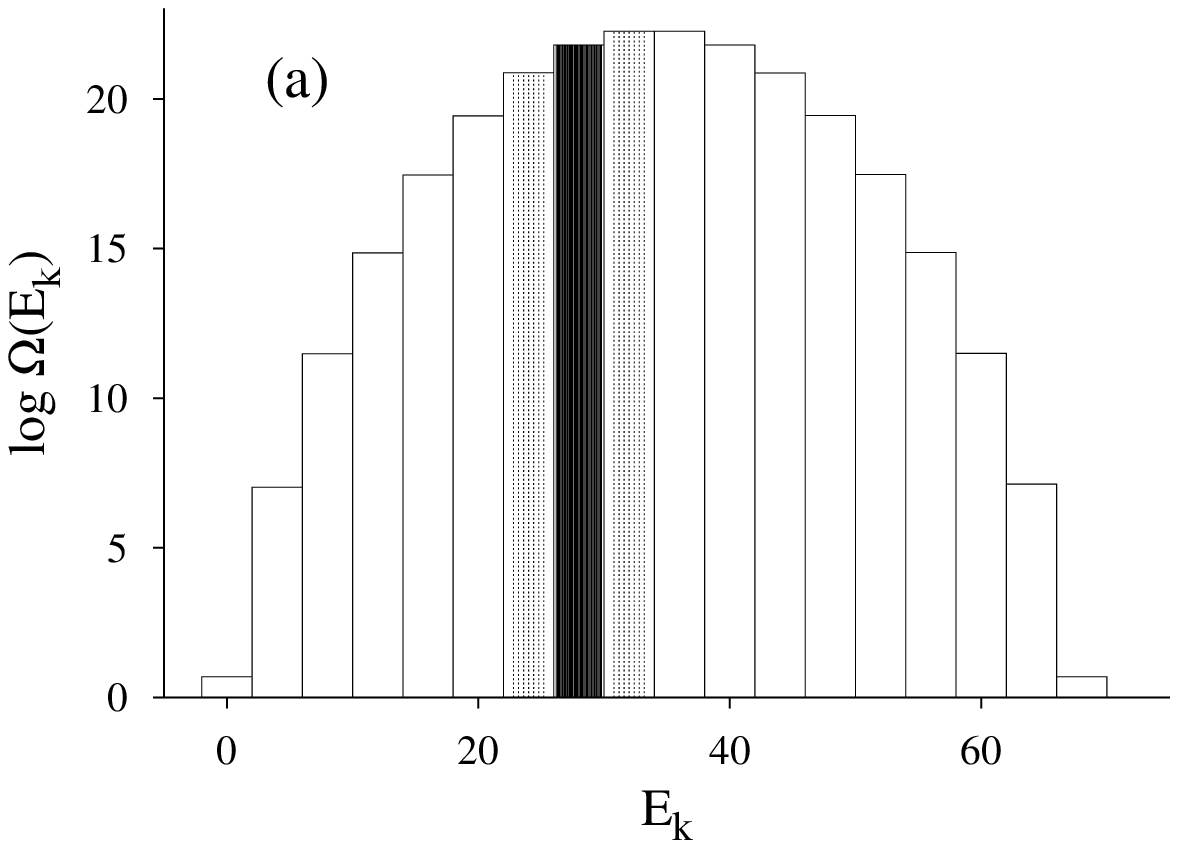}}
\centerline {\epsfxsize=10cm \epsfbox{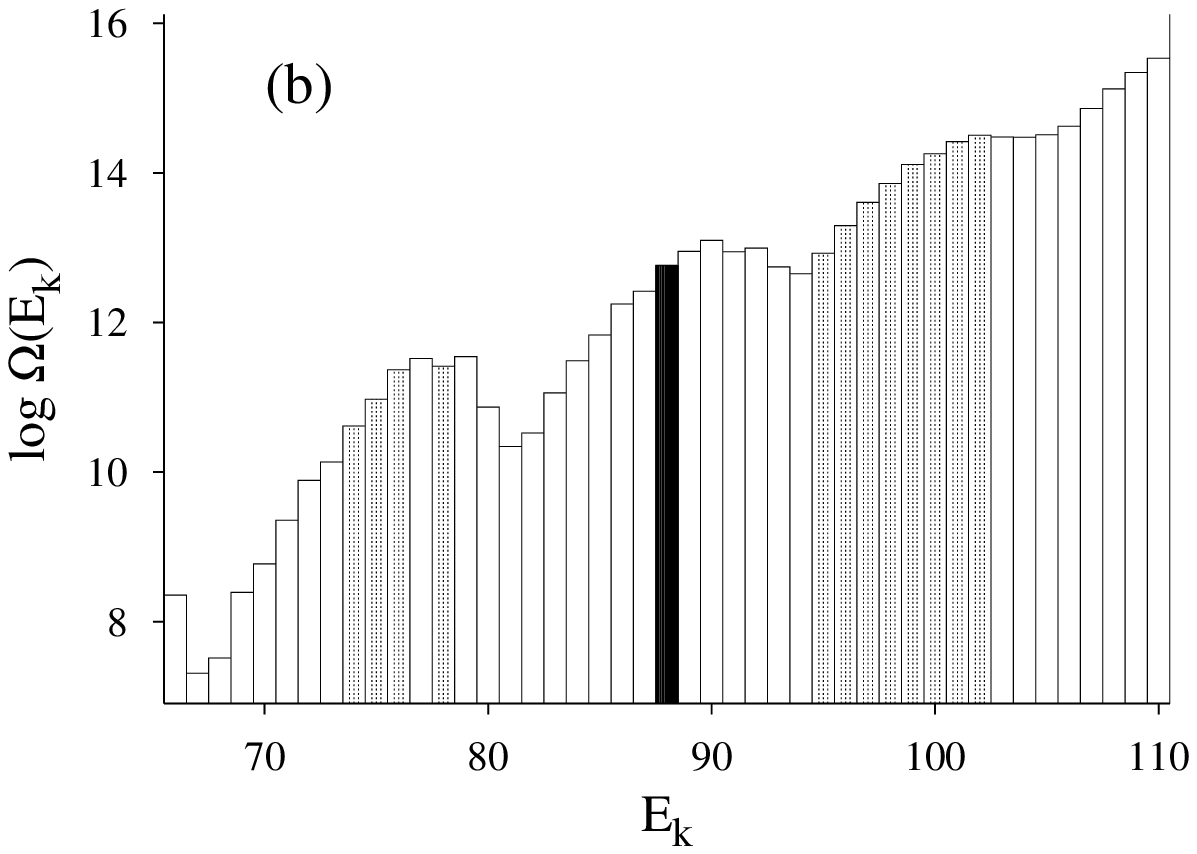}}
\caption{All the possible jumps from the energy level mark in black
generated by a single spin flip. (a) SRIM, (b) LRIM. Note in (b) the small 
modulation in $\Omega(E_k)$ of the order of $l_{E}=2  \sum_{i=1}^N
r_{ij}^{-\alpha}$ (See the text). 
\label{f4}}
\end{figure}

The same basic idea has been used in other short--range Hamiltonians
\cite{bha87,bha87b,bha87c,bha87d,sal99}. We are concerned now with the
extension of this method to consider long--range interactions, in
particular the LRIM introduced before. A modification needed is that the
energy values $E_k$ will represent now a continuum set of energies with a
bin size $\delta E$. The energy levels are then $E_k=k \delta E$
$k=1,\dots,M$ and $\Omega(E_k)$ counts all the configurations $i$ whose
energy $\varepsilon_i$ satisfies $E_k\le \varepsilon_i <E_k+\delta E$. In
other words, one makes the approximation of considering that all the
energies lying in one bin count as one single level. This turns out not to
be a critical point, although one has to check that the results are
independent, within the simulation errors, of the magnitude of $\delta E$. 

\begin{figure}[!ht]  
\centerline {\epsfxsize=10cm \epsfbox{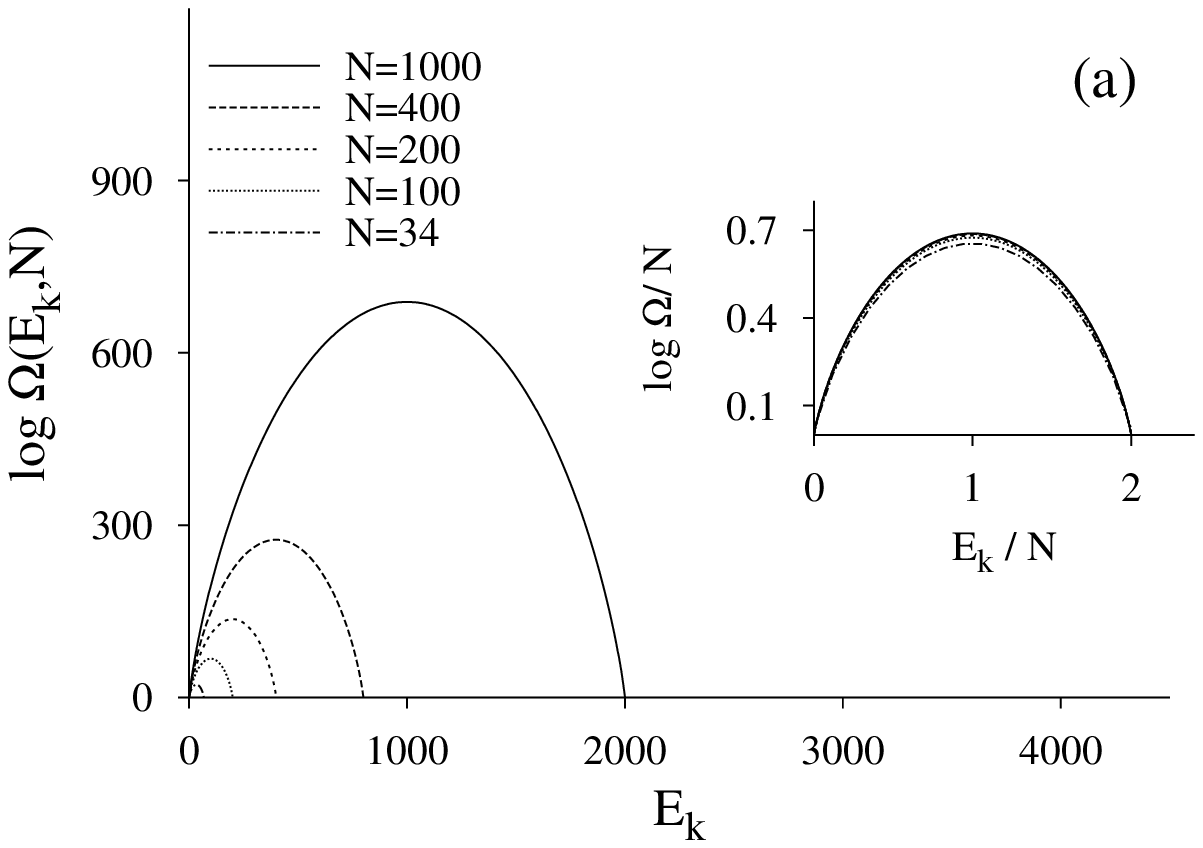}}
\centerline {\epsfxsize=10cm \epsfbox{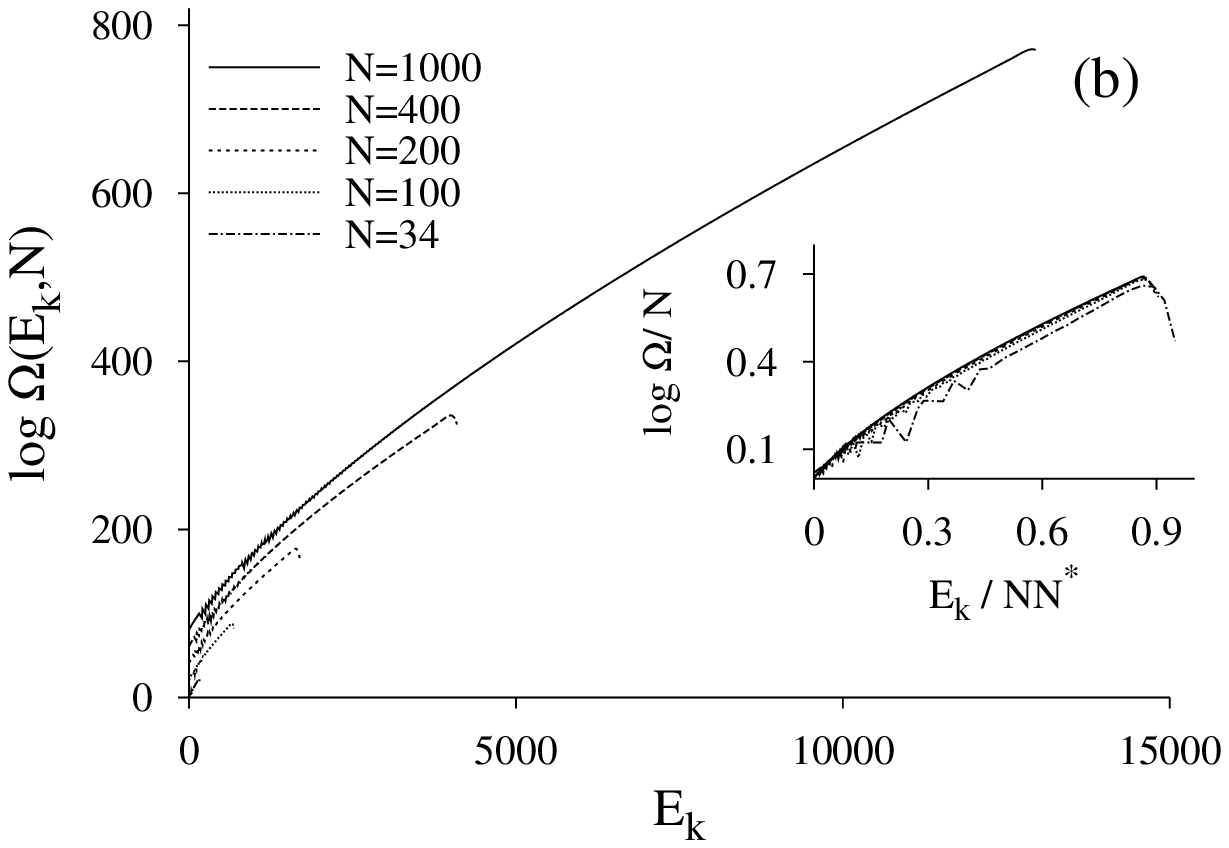}}
\caption{The number of states $\Omega(E_k)$ calculated using the HOW 
method for the 1--d Ising models for several values of $N$. (a) SRIM, (b)
LRIM. In (b) we have shifted vertically the curves to avoid overlapping.
The inserts show the collapse of all the curves using Eq. (\ref{e23}). 
\label{f5}}
\end{figure}

A more important point concerns the optimal size $l$ of the energy window
$\{E_k,E_{k+1},\dots,E_{k+l}\}$. Since, for a long--range interacting
system, a single spin flip can produce a very large change in the energy,
it is important not to choose $l$ too small. To make this point clear, we 
have calculated exactly the number of states  $\Omega(E_k)$ for a system
with $N=34$  by using the complete enumeration of the $W=2^{34}$ possibles
configurations, see Fig. \ref{f3}. Using these exact results we study the
energy changes that a single spin flip makes both in the SRIM and the LRIM
cases. A typical situation is shown in Fig.\ref{f4}. In this figure we plot
the histogram of the (exact) number of configurations using a value for the
bin $\delta E=4$ for the SRIM, and $\delta E=1$ for the LRIM. We select
several configurations belonging to one of the energy bins (marked black in
the figure) and we dash all the levels that are obtained from these
configurations by flipping one spin. We see that, as expected, the change
in energy for the SRIM brings the system to one of its neighboring energy
levels. However, for the LRIM the energy changes are very large and, in
fact, the nearest--neighbor energy levels can not be reached by using the
spin flip dynamics \cite{com5}. A measure of the typical change in energy
obtained when flipping one spin is estimated by considering the
ferromagnetic ground state configuration with all the spins pointing in the
same direction. One spin flip in this configuration produces a change in
energy $\Delta E=2 \sum_{i=1}^N r_{ij}^{-\alpha}$. The equivalent number of
energy bins is $\xi=\Delta E/\delta E$. We finally take the size of the
energy windows $l=3\xi$. In order to make sure that ergodicity is
satisfied, we adopt a large overlap of size $2\xi$ between the windows.
This means that a window goes from $E_k$ to $E_{k+3\xi}$, but the next
window goes from $E_{k+\xi}$ to $E_{k+4\xi}$ and so on. For the window
$\{E_k,E_{k+1},\dots,E_{k+3\xi}\}$ only those values in the interval
$\{E_{k+\xi},\dots,E_{k+2\xi}\}$  are considered for the evaluation of the
ratios $\Omega(E_k)/\Omega(E_n)$. To summarize, for the LRIM, it is
necessary that both the window size and the overlap between windows has the
correct size, depending on $\alpha$. For $\alpha=0.8$, used in our
simulations, we take $\xi=4$ independently of the system size and adjust
$\delta E$ accordingly. We have checked that $\xi=10$ gives the same
results within the numerical errors.  The number of configurations
necessary increases with the required accuracy. We have adopted in our
simulations the criterion that the minimum number of counts for any energy
bin within a window is $100$. The knowledge of the exact degeneration of
the ground state $\Omega(E_0)=2$ allows finally the calculation of all the
$\Omega(E_k)$.

\begin{figure}[!ht]  
\centerline {\epsfxsize=10cm \epsfbox{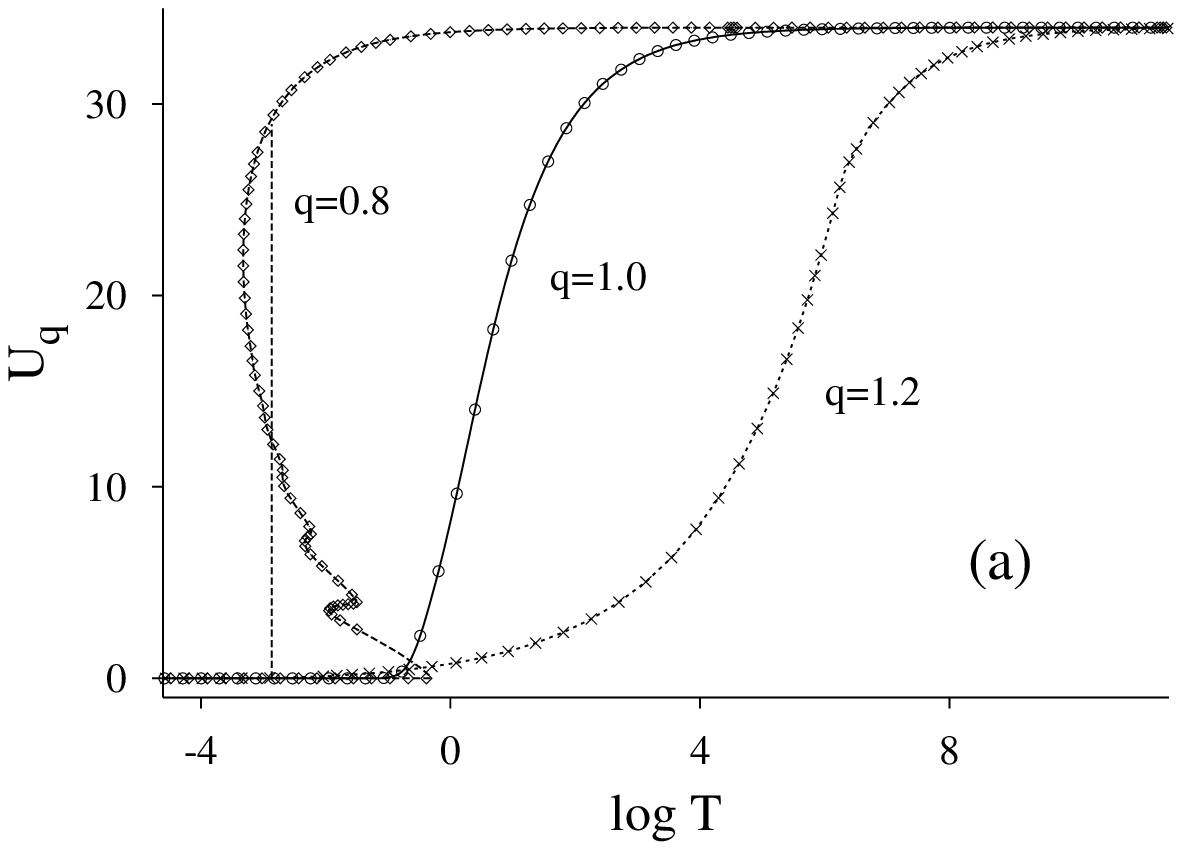}}
\centerline {\epsfxsize=10cm \epsfbox{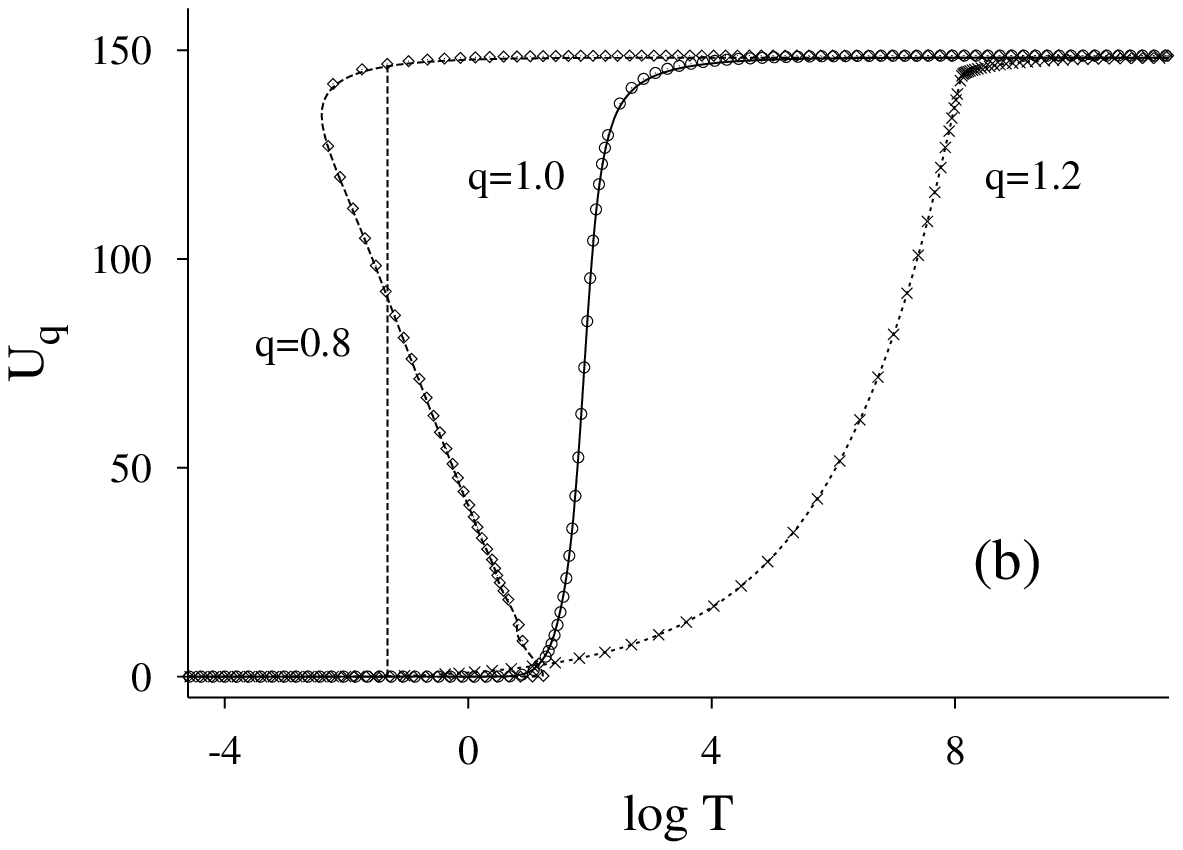}}
\caption{Internal energy $U_q(T)$ for 1--dimensional Ising models with
$N=34$ and $q=0.8,1.0,1.2$. Symbols are obtained using the HOW method and
lines show the exact results.  (a) SRIM, (b) LRIM.
\label{f6}}
\end{figure}

\begin{figure}[!ht]  
\centerline {\epsfxsize=10cm \epsfbox{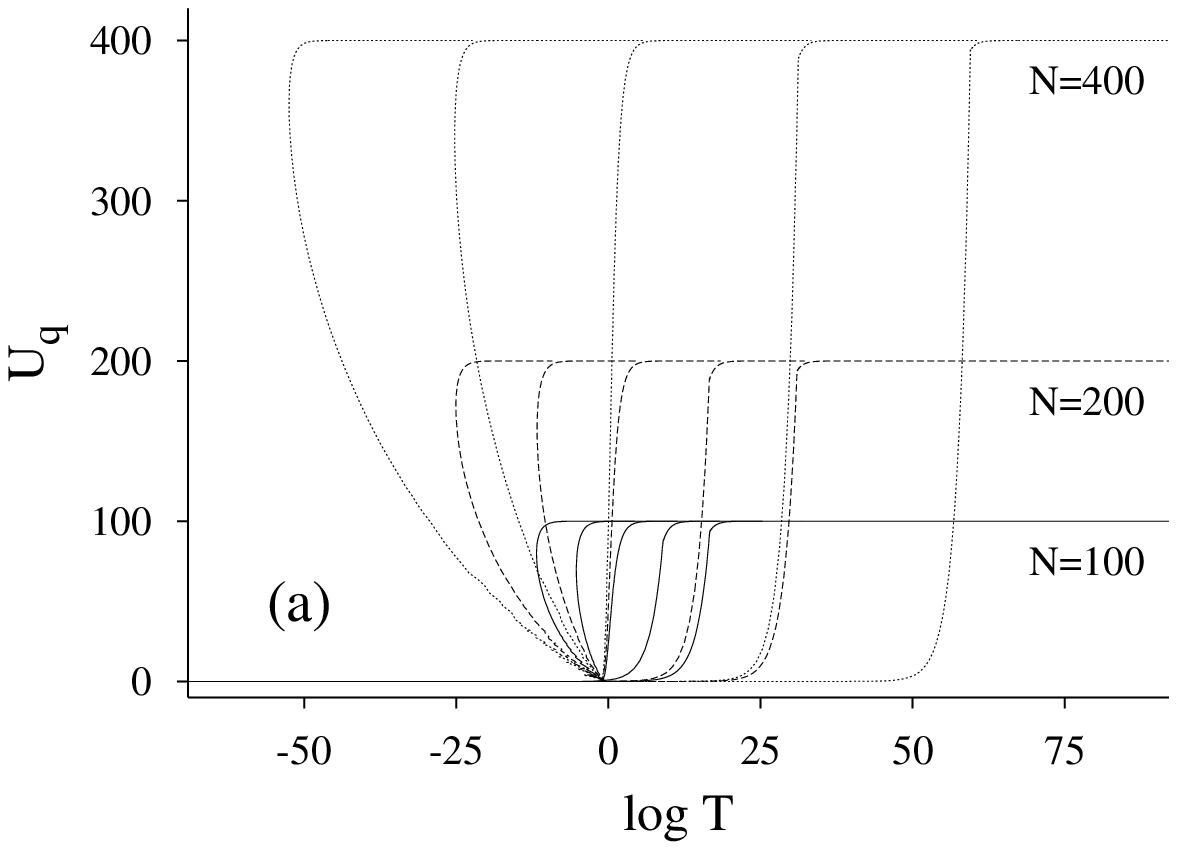}}
\centerline {\epsfxsize=10cm \epsfbox{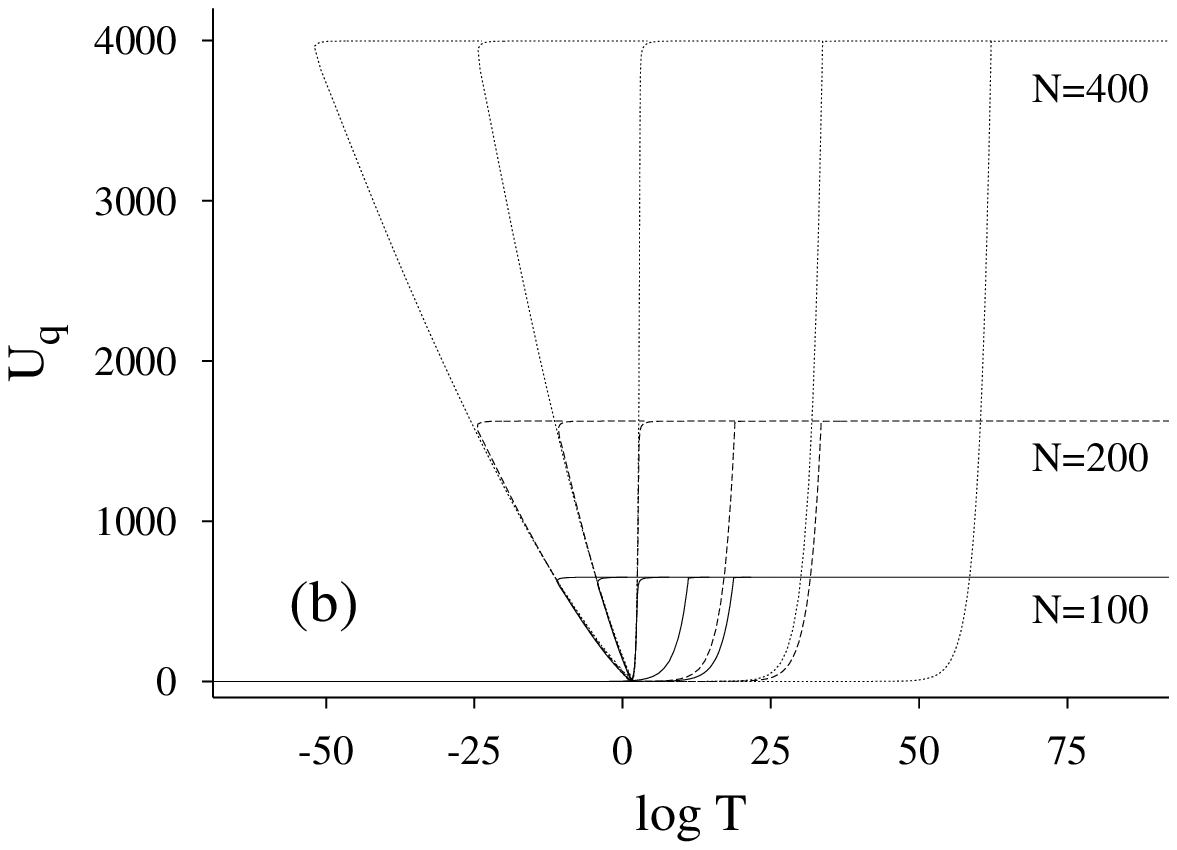}}
\caption{Internal energy $U_q$ as a function of temperature $T$ for
one--dimensional Ising models and different values of $N$ and $q$, as
coming from the application of the HOW method in the (a) SRIM and (b) LRIM.
The values of $q$ are $q=0.8,0.9,1.0,1.1,1.2$ (curves from left to right).
\label{f7}}
\end{figure}

In Fig. \ref{f3}, we plot the number of states computed either exactly or
by using the HOW method, for $N=34$ both for the SRIM and the LRIM. At the
resolution of the figure, the exact results and the approximate ones are
indistinguishable. This serves as a test that the HOW method, as
implemented here, is capable or reproducing accurately the number of states
in a known case. In Figs. \ref{f5}, we show the number of states for sizes
$N=34,100,200,400,1000$ as computed using the HOW method.  The inserts of
these figures show that this function scales as: 

\begin{equation}\label{e23} 
\Omega(E_k) = \e^{N \phi(E_k/N\tilde N)}
\end{equation} 

This is valid both for the SRIM and for the LRIM, if we recall that 
$\tilde N=1$ for the SRIM. The scaling function $\phi(x)$ is different for
the SRIM and the LRIM. For the SRIM, the result (\ref{eq:24}) leads to:

\begin{equation}\label{phi_srim}
\phi_{SRIM}(x)=\frac{x}{2}\ln\left(\frac{2}{x}-1\right)
-\ln\left(1-\frac{x}{2}\right)
\end{equation}

No analytical expression is available for the LRIM. Fig.  \ref{f6} compares
the result for the internal energy $U_q$ obtained using the HOW method and
the exact results obtained from the exact enumeration algorithm in the case
$N=34$.  Again, we can see that differences are too small to show up in
this plot.  Finally, in Fig.  \ref{f7} we make a similar plot for larger
values of $N$ obtained in this case by application of the HOW method.

\section{The Monte Carlo method}\label{s3} 

We have mentioned already that the usual Monte Carlo algorithms of the
Metropolis type can not be applied to Tsallis statistics, because the
probabilities $\{P_i\}$ are not known as a function of the temperature $T$.
However, it is possible to use them to compute averages as a function of
the parameter $T'$ because the probabilities are known as a function of
$T'$ except for a normalizing factor, which is irrelevant in the Monte
Carlo methods. Those averages can be performed by using the Metropolis
algorithm to generate configurations distributed according to the
probabilities $\{P_i\}$ as follows: consider the configuration $i$ with
energy $\varepsilon_i$, flip a spin chosen at random to produce
configuration $j$ with energy $\varepsilon_j$, accept this change with a
probability $\min (1,P_j/P_i)$. For the Boltzmann--Gibbs statistics the
acceptance probability is the celebrated factor
$\min[1,\exp(-\beta(\varepsilon_j-\varepsilon_i))]$. For Tsallis
statistics, Eq. (\ref{e9}), it is instead:  

\begin{equation}\label{e17}
P(i \to j)=\left \{ \begin{array}{lr}
0, & 1- (1-q) \beta' \varepsilon_j <  0\\
\min \Biggl[1,\frac {1-(1-q)\beta'  \varepsilon_j}{1-(1-q)\beta'
\varepsilon_i}\Biggr]^{\frac q {1-q}}, & {\rm otherwise}
\end{array} \right.\\  
\end{equation} 

This acceptance probability was used for the first time by I. Andricioaei
{\it et. al.} \cite{ioa96}. After generating ${\cal N}$ configurations, the
averages $\langle O \rangle_q$ at fixed $T'$ are obtained as the sum:
$\langle O\rangle_q= {\cal N}^{-1}\sum_{s=1}^{\cal N} O(s)$, where $O(s)$ 
is the value of the observable $O$ at the configuration $s$ in the sequence
of configurations generated by the Monte Carlo method.

\begin{figure}[!ht]  
\centerline {\epsfxsize=10cm \epsfbox{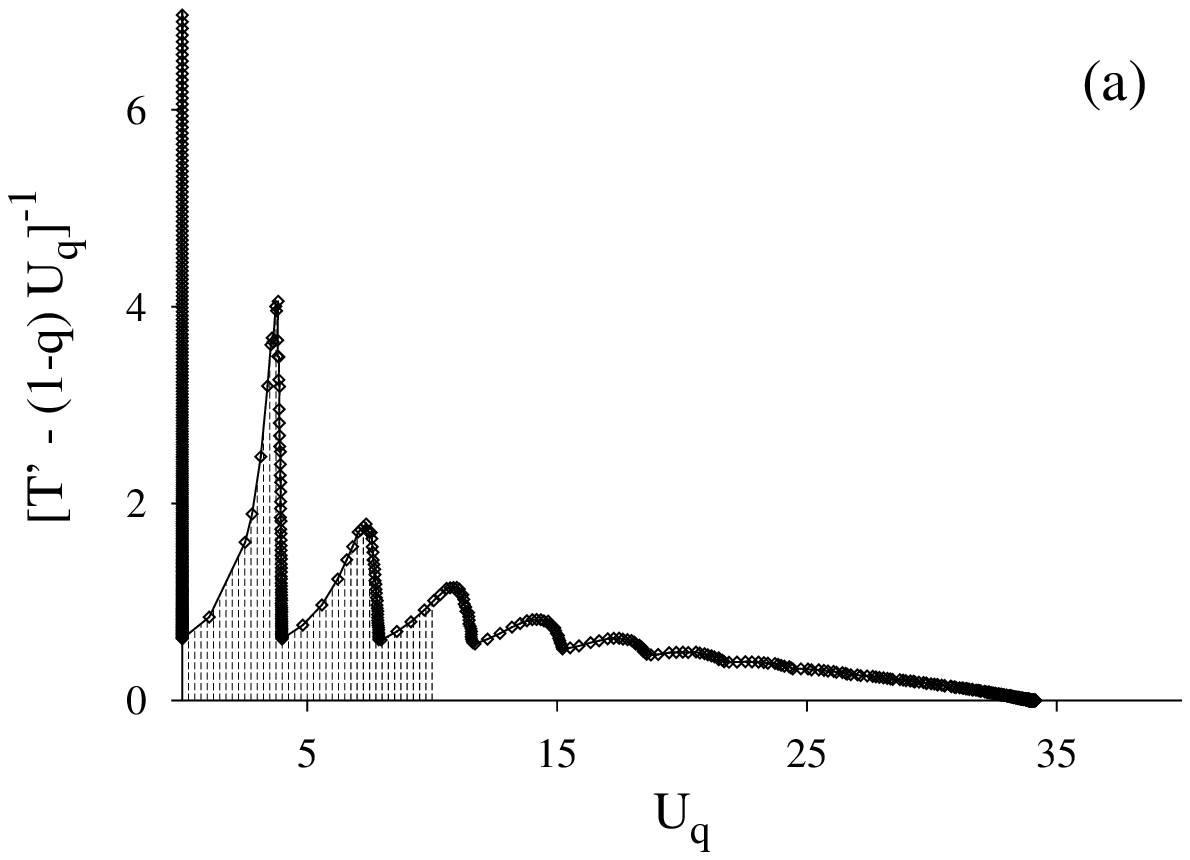}}
\centerline {\epsfxsize=10cm \epsfbox{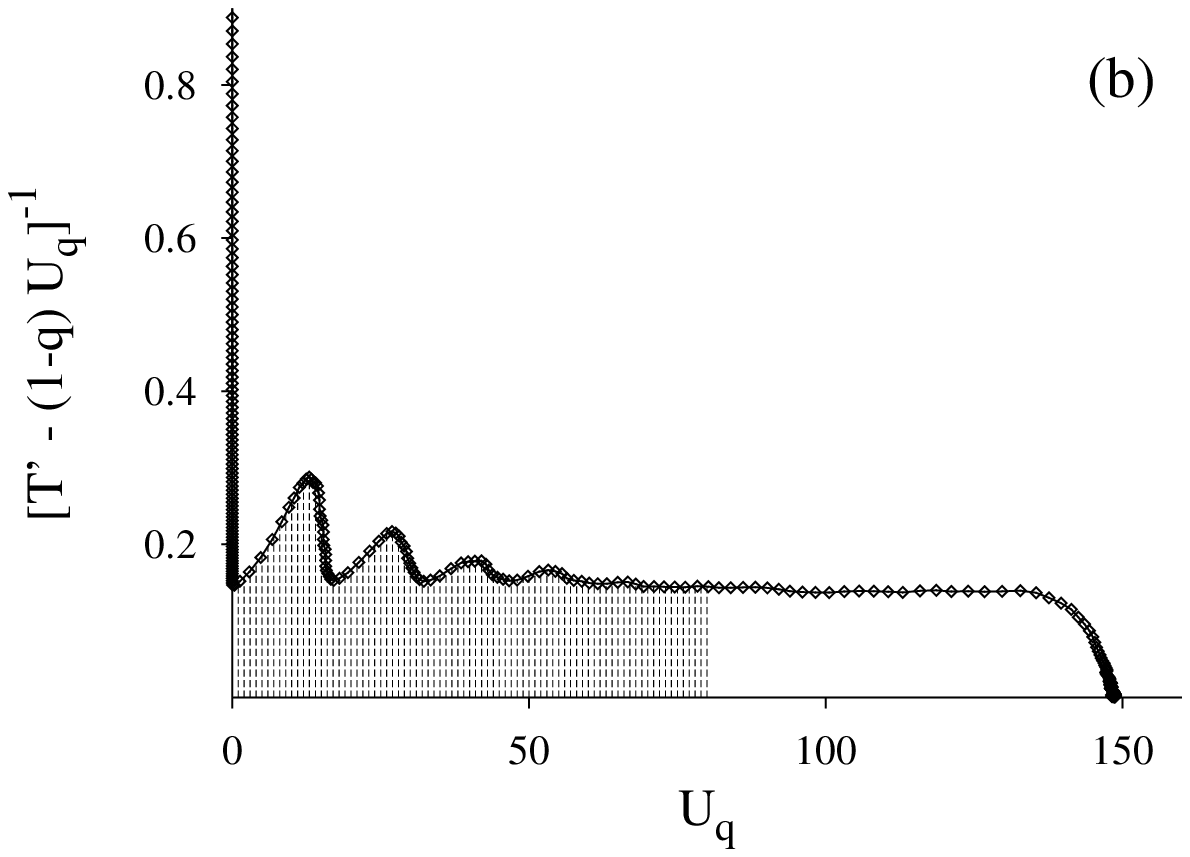}}
\caption{The integration procedure implicit in the Eq. (\ref{e20}) to
perform the  $T' \to T$ transformation in the Monte Carlo method. We use
here $N=34$, $q=0.6$ and $T^A=0$. (a) SRIM, (b) LRIM. 
\label{f8}}
\end{figure}

Still, we need to perform the $T'\to T$ transformation, in order to plot
averages with respect to $T$, using Eq. (\ref{e11}). In this equation, we
can use the Monte Carlo averages for the internal energy $U_q=\langle {\cal
H} \rangle$, but the entropy is yet unknown. In order to compute the
entropy, we combine Eq. (\ref{e11}) and  Eq. (\ref{e8}):

\begin{equation}\label{e18} 
\frac{\partial S_q}{\partial U_q} = \frac {1 + (1-q) S_q}{T' - (1-q) U_q} 
\end{equation} 

which can be integrated between arbitrary points $A$ and $B$:
 
\begin{equation} \label{e19}
\frac 1 {1-q} \ln [1+(1-q)S_q] \biggr|_A^B = \int\limits_{U_q(A)}^{U_q(B)} 
\frac {d U_q}{T' - (1-q) U_q} 
\end{equation} 

to obtain:

\begin{equation}\label{e20} 
S_q(B) = \frac {[1+(1-q)S_q(A)] 
\e^{\Bigl[ (1-q) \int\limits_{U_q(A)}^{U_q(B)}  
\frac{d U_q}{T' - (1-q) U_q}\Bigr]} -1} {1-q} 
\end{equation} 

\begin{figure}[!ht]  
\centerline {\epsfxsize=10cm \epsfbox{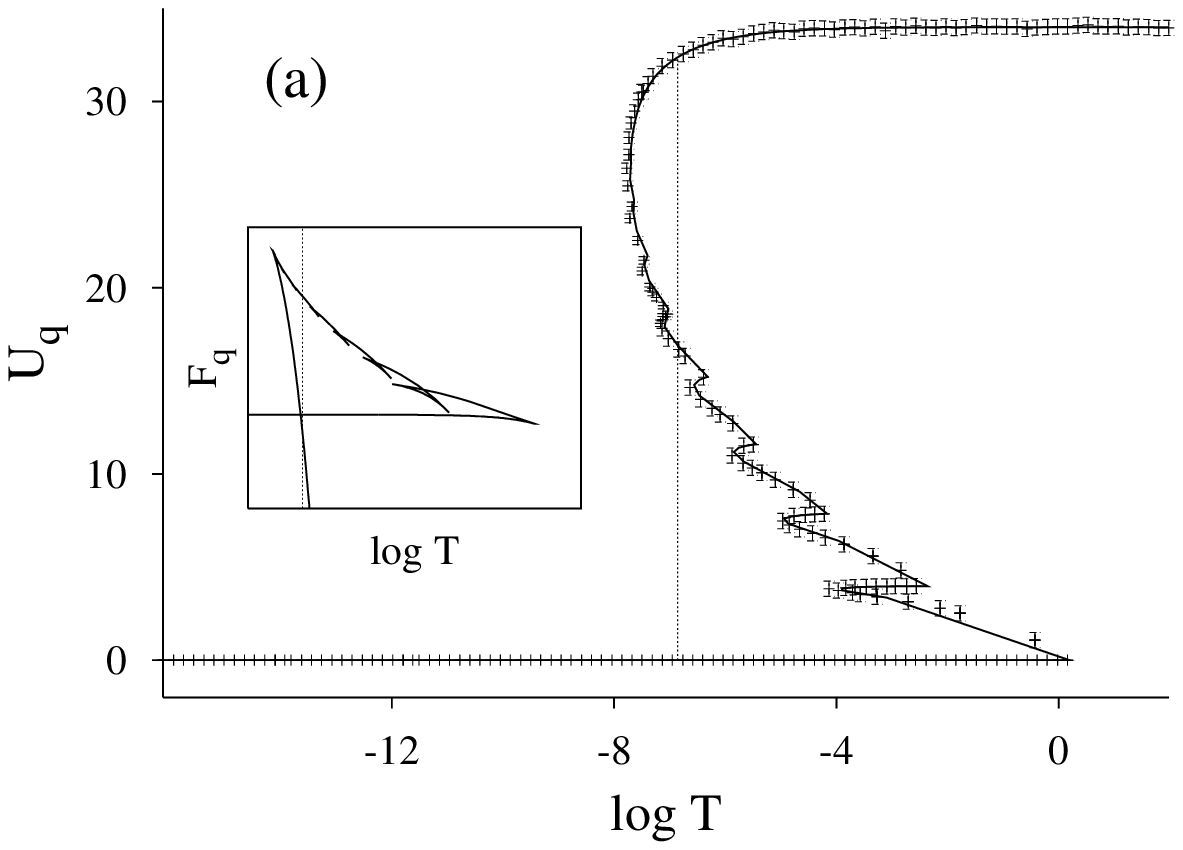}}
\centerline {\epsfxsize=10cm \epsfbox{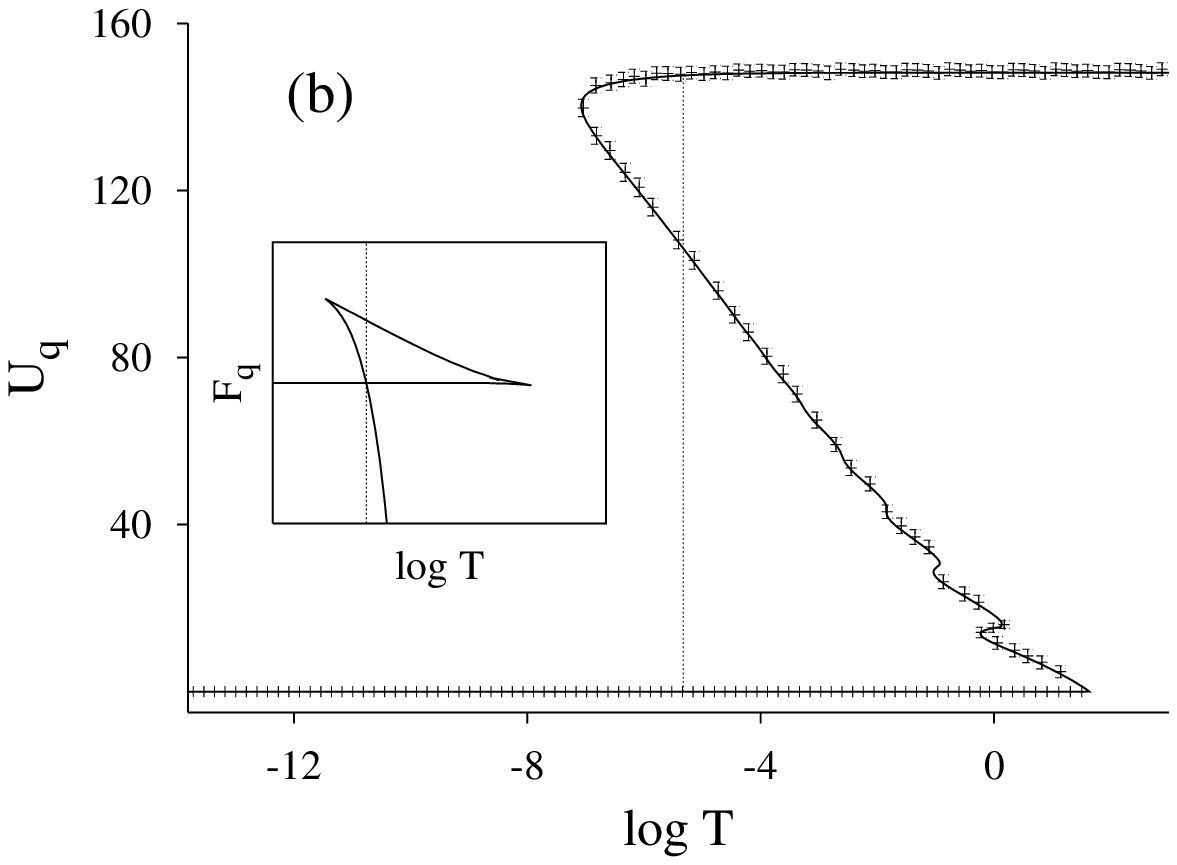}}
\caption{Internal energy $U_q$ as a function of the temperature $T$, for
$N=34$ and $q=0.6$. By symbols we show the results of the Monte Carlo
method and by lines the exact results. In the inserts we show the free
energy criterion explained in the main text. (a) SRIM, (b) LRIM.
\label{f9}}
\end{figure}

Finally, we need to know the value of the entropy, $S_q(A)$ at the initial
integration point A. This depends on the particular system considered, but
usually the extreme temperature cases are known. For the Ising models (both
long--range and short--range), Eq. (\ref{e14}), the limits $T \to 0$ and $T
\to \infty$ are: 

\begin{itemize} 
\item[i.] $S_q(T=0)=\frac {2^{(1-q)}}{1-q}$. 
\item[ii.] $S_q(T=\infty)=\frac {2^{N(1-q)}}{1-q}$.
\end{itemize}

We have implemented this Monte Carlo method using a system size $N=34$.
Fig. \ref{f8} shows that the function that needs to be integrated in order
to perform the $T' \to T$ transformation is a smooth one. Fig. \ref{f9}
compares the internal energy obtained by this Monte Carlo method with the
exact results obtained by the exact enumeration procedure showing the
validity of this Monte Carlo scheme.

The main disadvantage of this Monte Carlo method is that one might need to
simulate a large range of values of $T'$  to be able to perform accurately
the integration needed for the $T' \to T$ transformation. However, the use
of extrapolation techniques, such as the multiple histogram
reweighting \cite{fer88,fer89}, which permit to extend the results of a
simulation at a value of the temperature to a continuum range of
temperatures, allows to reduce dramatically the number of simulation points
needed \cite{sal00}.

\section{The scaling functions}\label{s5}

For extensive systems, the internal energy per particle is just a function
of the temperature, $U(N,T)/N=u(T)$.  Clearly, by definition, this scaling
relation does not hold for non--extensive systems and there has been some
recent interest in finding the correct scaling laws that apply to
non--extensive systems.  A first result was to obtain the scaling laws that
follow from the application of Boltzmann--Gibbs statistics to a genuinely
non--extensive system such as the LRIM in the regime $\alpha < d$.  The
results for the internal energy, $U$, the magnetization $M$ (defined as
$M=|\sum_{i=1}^N s_i|$ and computed using Eq. (\ref{e22c})), the Helmholtz
free energy $F$ and the entropy $S$ can be summarized by the following
relations \cite{tsa95b,can96,jun95,sam97,gri96,reg99}:

\begin{eqnarray}
U(N,T) & = & N \tilde N  u(T/ \tilde N)\label{e29a}\\ 
M(N,T) & = & N  m(T/\tilde N) \label{e29b}\\ 
F(N,T) & = & N \tilde N  f(T/\tilde N) \label{e29c}\\ 
S(N,T) & = & N s(T/\tilde N)\label{e29d}
\end{eqnarray}

where $m$,$u$,$f$,$s$ are the scaling functions. The argument justifying
these scaling laws can be summarized as  follows:  the internal energy and
the entropy appear in the definition of the free energy as $F=U-TS$,
therefore one expects that $U$ and $TS$ should have the same behavior for
large $N$. Since $U$ scales as $N \tilde N$ and $S$ scales as $N$ one
obtains that $T$  must scale as $\tilde N$ thus leading to the previous
scaling ansatzs. Note that the SRIM case is recovered from the LRIM case in
the limit  $\alpha \to \infty$, when $\tilde N \to 1$ and the scaling
relations, Eq. (\ref{e29a}-\ref{e29d}), become the standard ones for
extensive systems.

We present now the extensions of these scaling laws in the case that the
models are considered under the rules of Tsallis entropy \cite{sal99}. In
the case $q\ne 1$, the entropy is no longer an extensive quantity (this is
true both for the SRIM and the LRIM). In order to generalize the argument
of the previous paragraph giving the correct scale factor for the
temperature, we derive from Eq. (\ref{e2}) the following general expression
for the entropy $A_q(N)$ of a set of $N$ {\sl independent} particles:

\begin{equation}\label{e30} 
A_q(N) =S_q(N,T=\infty) = \frac {[1+(1-q) S_q(1)]^N-1}{1-q} 
\end{equation} 

here $S_q(1)$ is the one particle entropy. In the Ising model case, one
particle can be in any of the two states with  probability $1/2$, yielding:
$S_q(1)=[1-2(1/2)^q]/(q-1)=[2^{1-q}-1]/(1-q)$. After replacing in Eq.
(\ref{e30}), we obtain: 

\begin{equation}\label{e31}  
A_q(N) = \frac {2^{N(1-q)}-1}{1-q} 
\end{equation} 

Assuming that Tsallis entropy will be scaled generically with $A_q(N)$, we
now assume that $U$ and $TS$ scale in the same way as $N\tilde N$.  Since
$TS/N\tilde N =[T A_q(N)/N\tilde N][S/A_q(N)]$ we conjecture that the
temperature has to be scaled with $N'\equiv N \tilde N/A_q(N)$.  However,
it turns out that the numerical results do not support this expression for
the rescaling factors in the case $q>1$.  Therefore, we write the scaling
relations in the following more general form:

\begin{eqnarray} 
U_q(N,T) & = & N \tilde N u_q(T/N_U') \label{e32a}\\  
M_q(N,T) & = & N m_q(T/N_U') \label{e32b}\\ 
F_q(N,T) & = & N \tilde N f_q(T/N')\label{e32c}\\ 
S_q(N,T) & = & A_q(N) s_q(T/N_S') \label{e32d}
\end{eqnarray} 

The previous argument, valid in the case $q\le 1$, implies simply
$N'_U=N'_S=N'$.  For consistency in the notation, we define $A_q^U(N)$ and
$A_q^S(N)$ by means of $N_U'\equiv N \tilde N/A^U_q(N)$ and $N_S'\equiv N
\tilde N/A_q^S(N)$, respectively.  Notice that for $q=1$ it is $A_1(N)
\propto N$ and the scaling laws (\ref{e29a}-\ref{e29d}) are recovered.  In
order to obtain a good scaling description in the case $q>1$ it is seen
numerically that one needs to assume the limits $A_q^U(N)\sim
2^{N(1-q)}/(q-1)$ and $A_q^S(N)\sim 2^{N(q-1)}/(q-1)$. A unifying
expression that reduces to the necessary ones for large $N$ and for all
values of $q$ is:

\begin{equation}\label{e33}
A_q^S(N)=\frac {2^{N(1-q)}}{q-1}, \hspace{2.0cm} 
A_q^U(N)=\frac {A_q(N)^2}{A_q^S(N)}
\end{equation}

Although we lack a satisfactory explanation for these relations, we note
that similar scaling factors have been used previously to plot in the same
scale curves for the specific heat in infinite--range Ising models and
non--interacting ideal paramagnet \cite{nob95,nob96}.   

\begin{figure}[!ht]  
\centerline {\epsfxsize=10cm \epsfbox{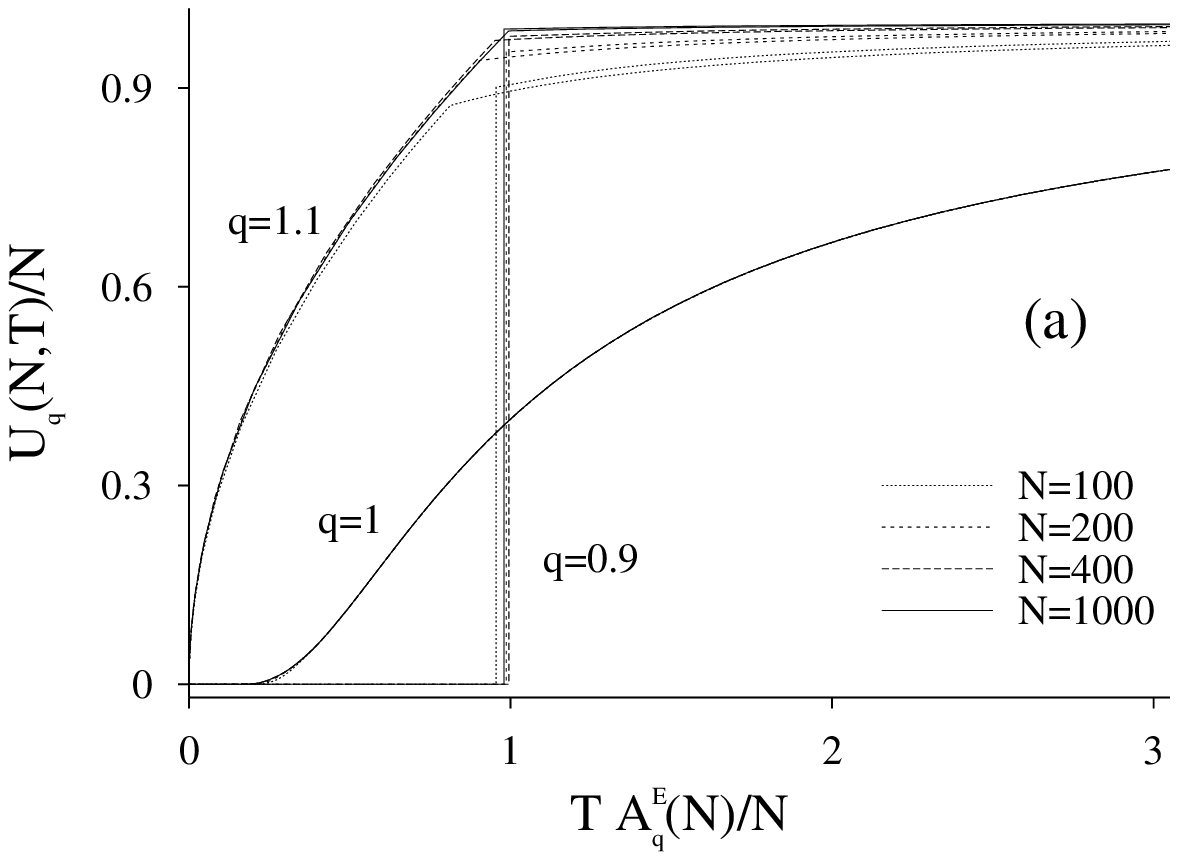}}
\centerline {\epsfxsize=10cm \epsfbox{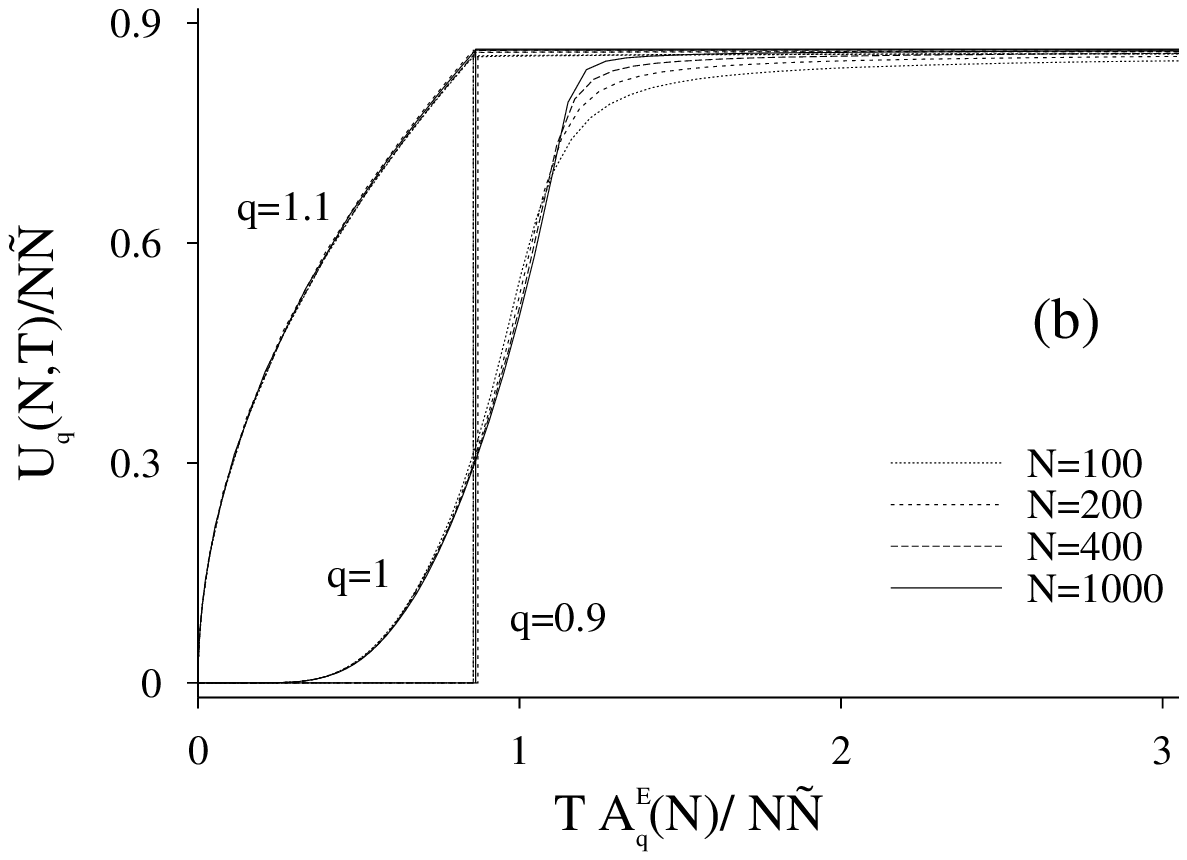}}
\caption{Internal energy $U_q(T)$ plotted in order to check the scaling
relation Eq.(\ref{e32a}) for $q=0.9,1.0,1.1$ and different system sizes.
(a) SRIM, (b) LRIM. 
\label{f10}}
\end{figure}

\begin{figure}[!ht]  
\centerline {\epsfxsize=9.0cm \epsfbox{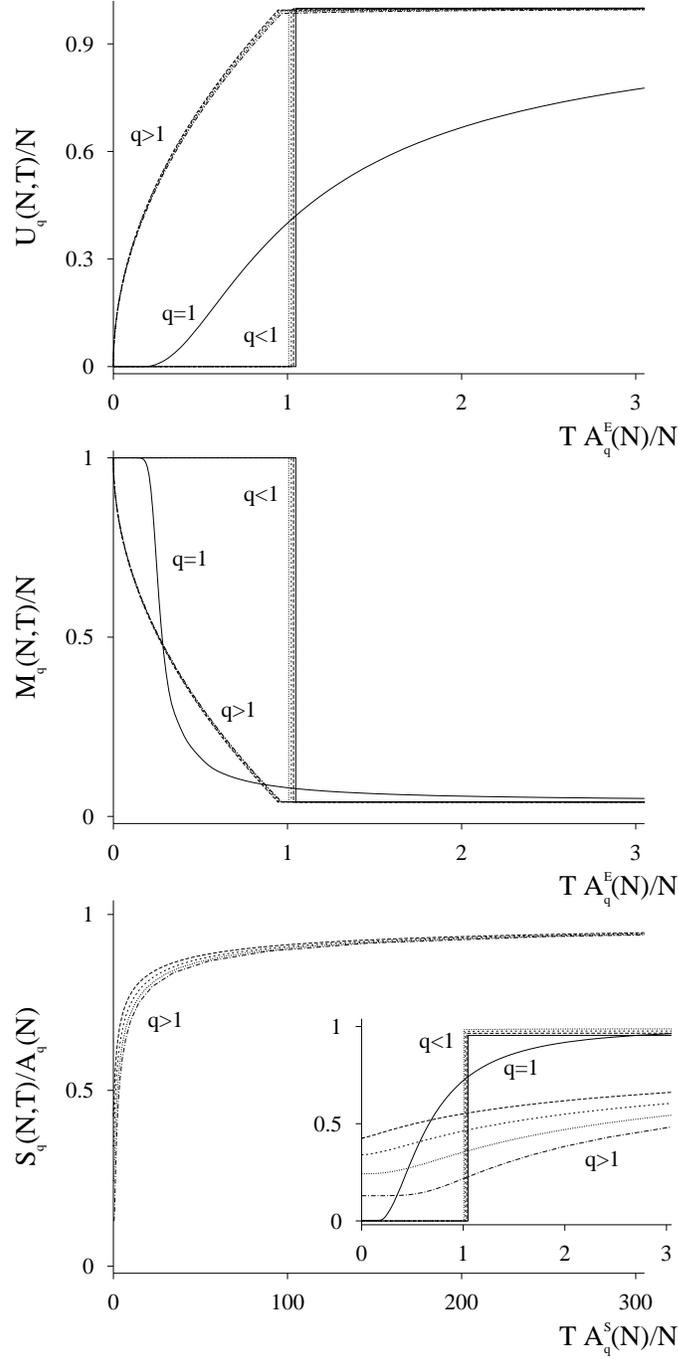}}
\caption{Internal energy (top graph), magnetization (middle graph) and
entropy (lower  graph) plotted in order to check the proposed scaling
relations Eqs. (\ref{e32a},\ref{e32b},\ref{e32d}) for the short--range
Ising model (SRIM). We have use $N=1000$ and the curves with $q<1$ include
$q=0.2,\,0.4,\,0.6,\,0.8$ while the curves with $q>1$ include
$q=1.2,\,1.4,\,1.6,\,1.8$. For clarity, in the  entropy plot, the insert
shows all the values of $q$, whereas the main  plot shows only $q>1$.
\label{f11}}
\end{figure}

\begin{figure}[!ht]  
\centerline {\epsfxsize=9.5cm \epsfbox{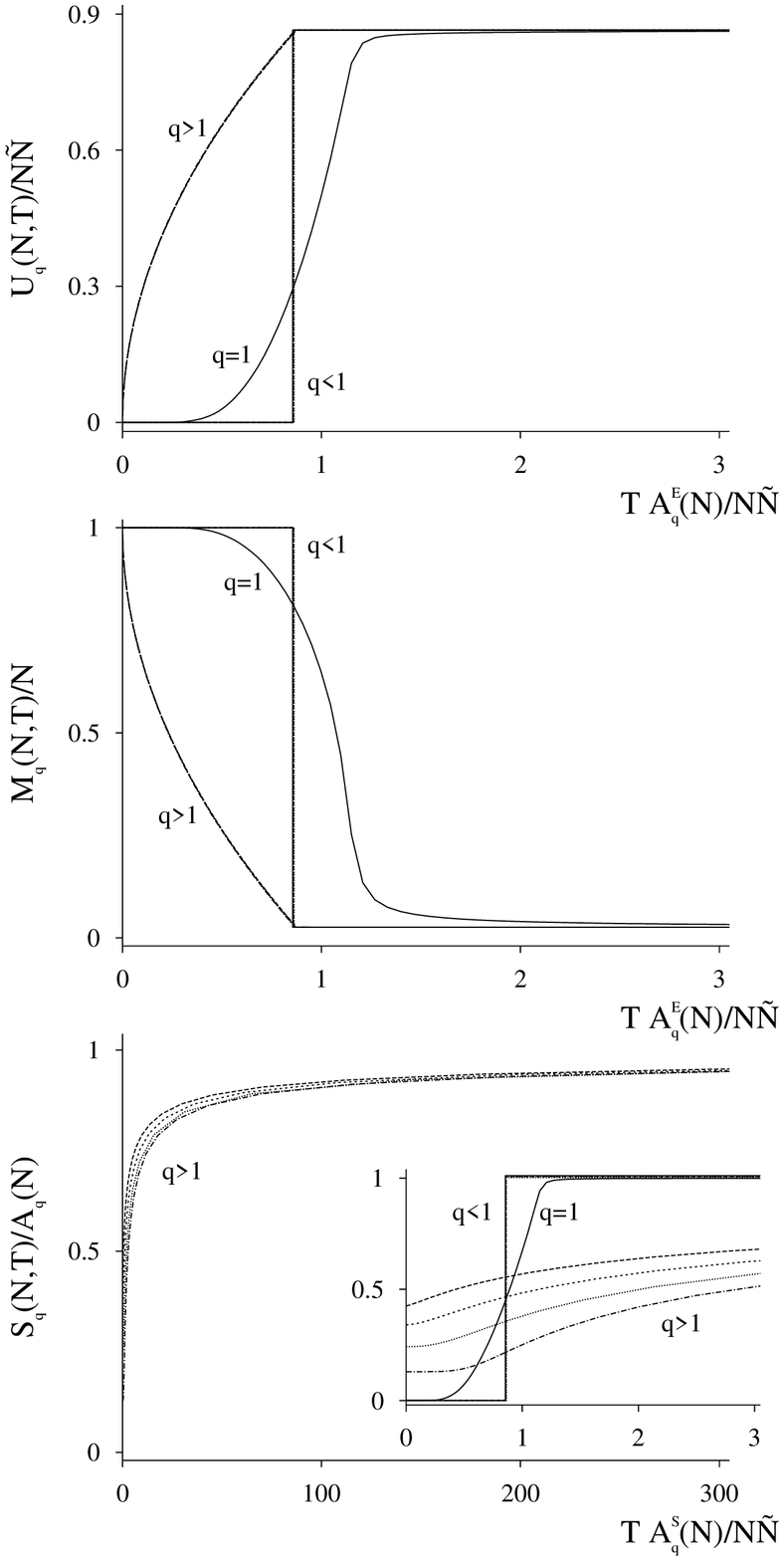}}
\caption{Same plot that Fig. \ref{f11} but for the long--range Ising model
(LRIM). We plot all the same $q$ values than in Fig. \ref{f11}, although
the different curves are almost indistinguishable with the  resolution of
this figure.
\label{f12}}
\end{figure}

In order to check the validity of these scaling relations, we have used the
HOW method to simulate the one-dimensional SRIM and LRIM with $\alpha=0.8$,
for system sizes $N=34,100,200,400,800,1000$, and several values for the
non--extensive parameter $q \in [0.1,1.9]$.  We test the proposed scaling
relations, Eq.  (\ref{e32a}-\ref{e32d}), by plotting the scaled results in
Figs.  \ref{f10},\ref{f11},\ref{f12}.  One can observe in the Fig.
\ref{f10} that, for the same value of $q$, the collapse of curves of
different sizes $N$. This is similar to what has been observed in the $q=1$
case \cite{can96}.

It is more remarkable the fact that, with the previous choice for the
scaling factors, all the $q<1$ data collapse in a single curve. The same
thing occurs for the $q>1$ curves. Therefore, data can be described by just
three universal scaling functions, corresponding to $q<1$, $q=1$, and $q>1$
regimes respectively (See Figs. \ref{f11},\ref{f12}).

One can see in Figs. \ref{f11},\ref{f12} that the collapse in the entropy
curves for $q>1$ is very poor. This is easily understood by noticing that
the low temperature limit of the entropy for infinite system size is
$S_q(T=0) =(1-2^{1-q})/ (q-1)$ whereas the high temperature limit is
$S_q(T\to\infty)=1/(q-1)$ and those two finite values can not be rescaled
simultaneously. This is different of what happens for the internal energy
and the magnetization for which the limits $T\to 0$ and $T\to \infty$
coincide for different values of $q$. Finally, the scaling for the free
energy follows directly from its definition $F_q=U_q-TS_q$. For $q \le 1$
it is  $f_q(x)=u_q(x)-xs_q(x)$, whereas for $q > 1$ and in the limit of
large $N$, the scaling function is given simply by $f_q(x)=u_q(0)-xs_q
(\infty)=-x$.

In summary, the scaling laws given by Eqs. (\ref{e32a}-\ref{e32d}) work for
all values of $q$ when using the scaling  factors given by Eqs.(\ref{e33}).
Moreover, the scaling functions $u_q$, $m_q$ and $f_q$ adopt only three
different forms for each magnitude corresponding to $q>1$, $q=1$ and $q<1$,
both in the SRIM and LRIM cases.

\section{Thermostatistics using standard mean values}\label{s6} 

It has been shown recently \cite{men97,pla97b,gue96} that the use of the
standard rule for the calculation of the mean values in Tsallis statistics
provides also a valid thermodynamical formalism. By ``standard" rule we
mean the use of the first option for the averages in which $u(p_i)=p_i$ is
used in (\ref{eq4}). Moreover, it has been argued \cite{tsa98,rag99} that
the results of using this first option coincide with the results of the
third option (the one used up to here in this paper) with a trivial change
in the parameter $q\to 1/q$. In this section we show that it is possible
indeed to map the results of one option into the results of the other,
although the relation between them implies, besides the previous change in
the parameter $q$, a non--trivial mapping for the temperature. Numerical
results using the techniques developed in the previous sections, will allow
us to plot the relation between the temperatures of the two options.

For the sake of clarity in the exposition we will use the subindexes ``1"
and ``3" to denote the results one obtains in each option. Hence, the 
first option seeks the maximization of

\begin{equation}\label{eb5} 
S_1(q)=\frac {1- \sum_{i} p_i^q}{q-1} 
\end{equation} 

subject to the canonical ensemble constrains: $p_i>0$, $\sum_i p_i=1$,
$\sum_i \varepsilon_i p_i=U$. The third option, on the other hand,  seeks
the maximization of

\begin{equation}\label{ea5} 
S_3(q)=\frac {1- (\sum_{i} p_i^{1/q})^{-q}}{q-1} 
\end{equation}

subject to the same constrains. Of course, in the third option, the
probabilities $p_i$ should be interpreted as ``escort" probabilities, but
this interpretation has no practical consequence whatsoever in the
calculation of the averages. The key point now is that both entropies are
related by:

\begin{equation}\label{eqs1s3}
S_1(1/q)=G_q[S_3(q)]
\end{equation}

\begin{figure}[!ht]  
\centerline {\epsfxsize=9.5cm \epsfbox{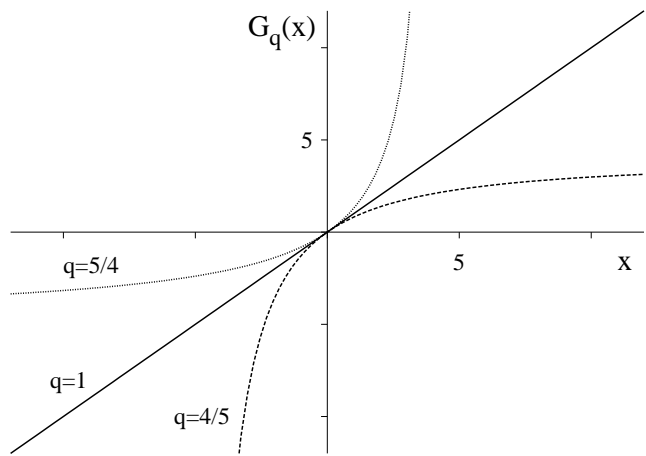}}
\caption{Transformation function between the two entropic forms $S_3(q)$
and $S_1(1/q)$ as indicated in Eq. (\ref{eqs1s3}).
\label{f13}}
\end{figure}

Where $G_q(x)=\frac{q}{1-q}[1-(1+(1-q)x)^{-1/q}]$ is a monotonically
increasing function of $x$. This function satisfies the property:
$G_q^{-1}(x) = G_{1/q}(x)$, see Fig.(\ref{f13}). Hence, the same set of
probabilities $\{p_i\}$ that maximize $S_3(q)$ for a given value of $U$ 
will maximize $S_1(1/q)$, for the same value for $U$. However, the fact
that the probabilities coincide in both options does not mean that the
averages computed using these probabilities coincide {\sl when they are
plotted as a function of the temperature} because it turns out that there
is a non--trivial relation between the temperatures of both options. Let us
denote by $T_1$ and $T_3$, respectively, the temperatures of the 1st and
3rd options. They can be defined as the (inverse of the)  Lagrange
multiplier needed to satisfy the constraint of fixed mean energy or,
alternatively, they have been shown to satisfy the relations 
\cite{pla97b}:

\begin{equation}
1/T_1(q)=\frac{\partial S_1(q)}{\partial U} \hspace{2.0cm}
1/T_3(q)=\frac{\partial S_3(q)}{\partial U}
\end{equation}

Using (\ref{eqs1s3}) we find the desired relation between the two
temperatures:

\begin{equation}
T_3(q)=T_1(1/q)/G'_{1/q}[S_1(1/q)]
\end{equation}

$G'_q(x)$ is the derivative of $G_q(x)$. After substitution of
Eq.(\ref{eb5}), we find:

\begin{equation}
\label{t1t3}
T_3(q)=T_1(1/q)(\sum_i p_i^{1/q})^{q+1}
\end{equation}

Therefore, it is true that the results of the third option at the value $q$
of the parameter can be obtained from those of the first one at the value
$1/q$. However, the mapping requires a non--trivial rescaling of the
temperature, as given by Eq.(\ref{t1t3}). Let us recall again that only the
dependence with $T$ does not vary when changing the zero of energies and,
hence, can be the only physically relevant one.

\begin{figure}[!ht]  
\centerline{\epsfxsize=9.5cm \epsfbox{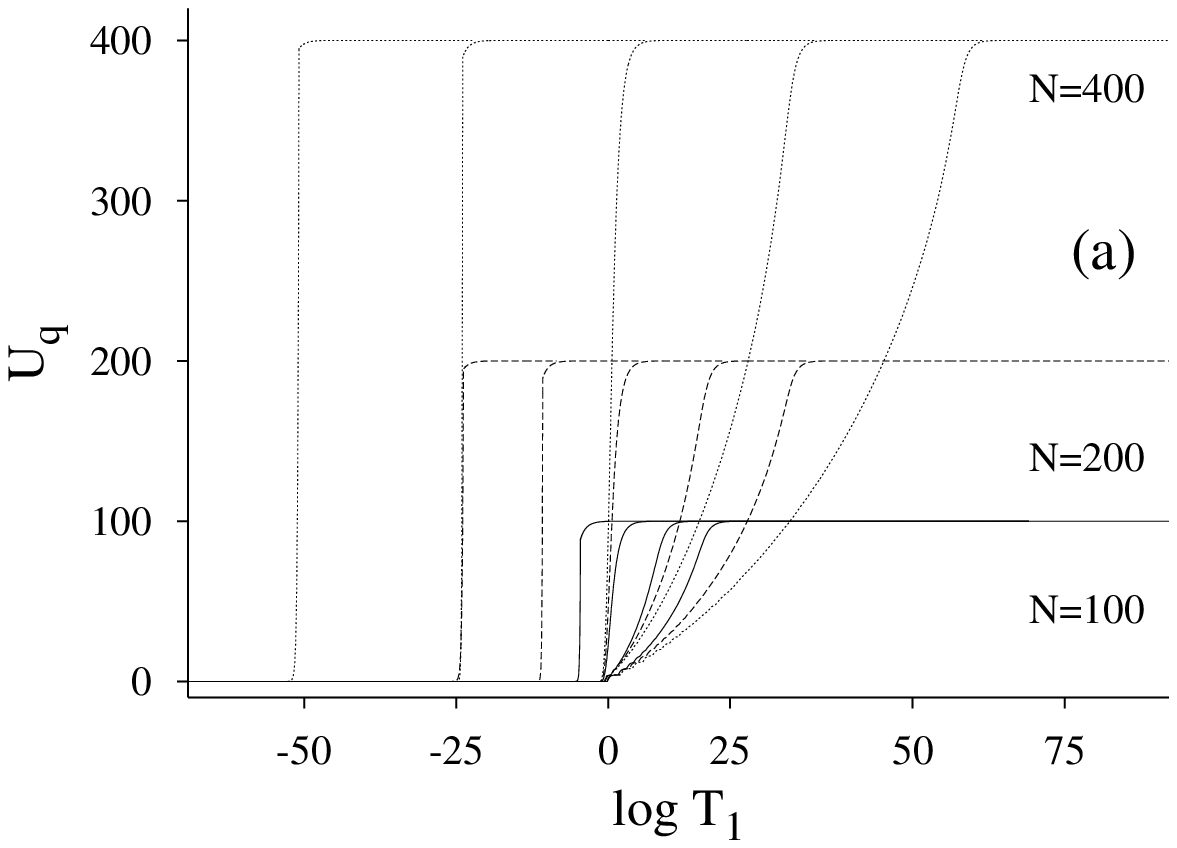}} 
\centerline{\epsfxsize=9.5cm \epsfbox{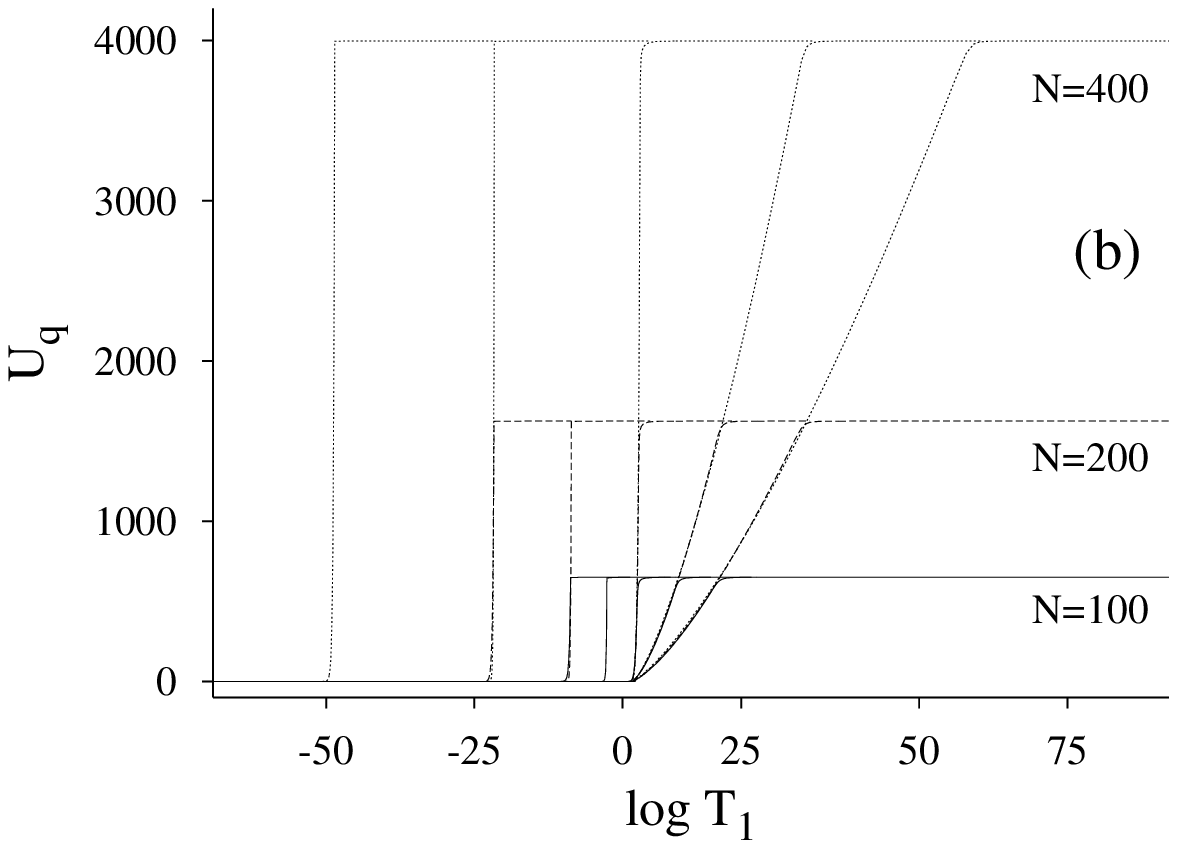}} 
\caption{The same plot as in Fig.  (\ref{f7}) but using lineal mean values
instead non--lineal ones for the values of $N$ indicated and
$q=0.8,0.9,1.0,1.1,1.2$.  (a) SRIM, (b) LRIM.
\label{f14}}
\end{figure}

\begin{figure}[!ht]  
\centerline {\epsfxsize=9.5cm \epsfbox{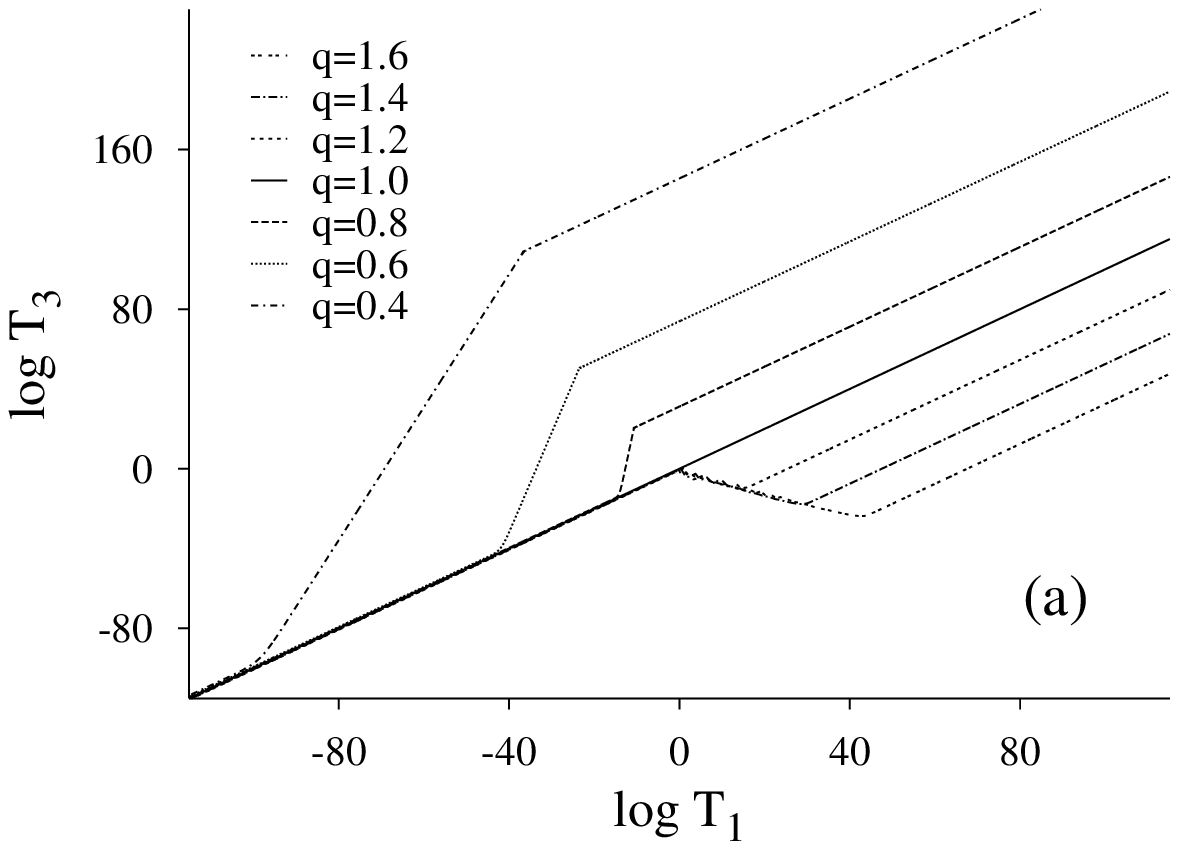}}
\centerline {\epsfxsize=9.5cm \epsfbox{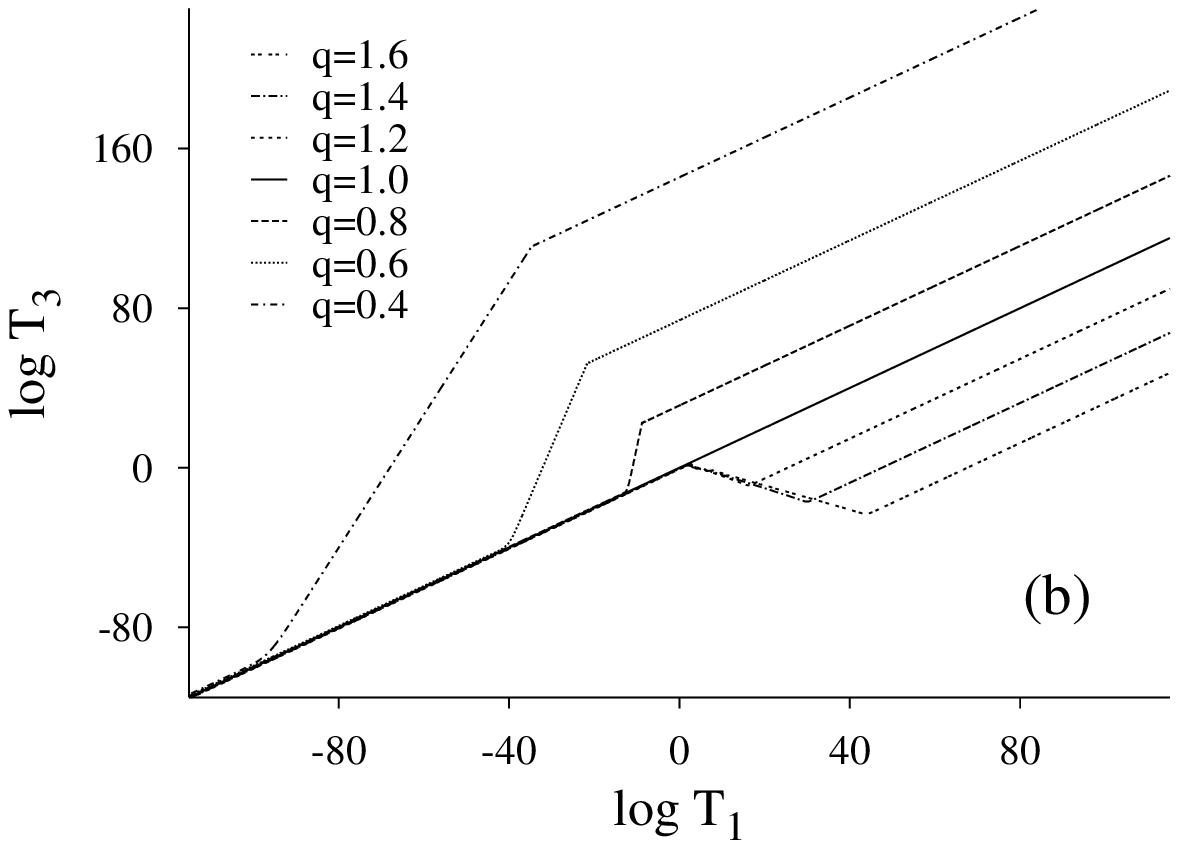}}
\caption{Transformation between temperatures $T_1$ and $T_3$ using Eq.
(\ref{t1t3}) for $N=100$ and the indicated $q$ values. (a) SRIM, (b) LRIM.
\label{f15}}
\end{figure}

In order to give an alternative explanation of the relation between the
temperatures of both options, let us write down the solutions for the
probabilities using the $\beta'$ parameter. For the third option, the
solution is read directly from Eq.(\ref{e9}) that we rewrite using the
notation of this section:

\begin{equation}\label{ea9}
p_i = \left \{ \begin{array}{lr}
0, & 1- (1-q) \beta_3' \varepsilon_i <  0\\
\frac {[1-(1-q) \beta_3' \varepsilon_i]^{\frac q {1-q}}}  
{\sum_{j} [1-(1-q) \beta_3' \varepsilon_j]^{\frac q {1-q}}}, & {\rm otherwise}
\end{array} \right.
\end{equation} 

where the parameter $\beta'_3=1/T'_3$ is related to the temperature $T_3$
by Eq.(\ref{e11}) which reads:

\begin{equation}\label{ea11} 
T'_3(q)= T_3(q)[1+(1-q) S_3(q)] + (1-q) U
\end{equation}

For the first option, it is possible to write the solution in the following
form:

\begin{equation}\label{eb9}
p_i = \left \{ \begin{array}{lr}
0, & 1- (1-1/q) \beta_1' \varepsilon_i <  0\\
\frac {[1-(1-1/q) \beta_1' \varepsilon_i]^{\frac 1 {q-1}}}  
{\sum_{j} [1-(1-1/q) \beta_1' \varepsilon_j]^{\frac 1 {q-1}}}, & {\rm otherwise}
\end{array} \right.
\end{equation} 

where the parameter $\beta'_1=1/T'_1$ is related to the temperature $T_1$
by \cite{com7}

\begin{equation}\label{eb11} 
T'_1(q)=  T_1(q)[1+(1-q)S_1(q)] +(1-1/q) U 
\end{equation}

Now, it is clear by comparing Eqs. (\ref{ea9}) and (\ref{eb9}) that the
probabilities $p_i$ of the third option computed at $q$ are equal to the
probabilities $p_i$ of the first option computed at $1/q$ provided that we
choose the same values for the primed temperatures, $T'_3(q)=T'_1(1/q)$.
After substitution of (\ref{ea11}) and (\ref{eb11}) and using (\ref{ea5}),
(\ref{eb5}) we recover Eq.(\ref{t1t3})

It is straightforward now to use the number of states $\Omega(E_k)$
obtained using the HOW method to compute
Eqs.(\ref{eb9},\ref{eb11},\ref{eb5}) by replacing the sums over the
configurations to sums over energy levels weighted by $\Omega(E_k)$. In
this way, we can perform the necessary averages implied in the first option
as well as the temperature transformation factor needed in Eq.
(\ref{t1t3}).  In Fig. \ref{f14} we plot the internal energy $U_q$ as a
function of the temperature using the standard averages of the first
option. The most noticeable difference with the results of the third
option, see figure (\ref{f5}) is that it is not necessary now to use the
Maxwell construction because there are no loops with the temperature. In
Fig. \ref{f15} we plot $T_3(q)$ vs. $T_1(1/q)$ in the LRIM and SRIM cases.
Using these two results, it is possible to obtain the averages within the
third option as a function of $T_3$. Of course, the results agree perfectly
with those shown in Fig. \ref{f7}. It is possible also to obtain from 
Eqs.(\ref{e32a}-\ref{e32d}) the scaling relations valid when the standard 
calculation of mean values is used for the calculation of thermodynamic
quantitities.

\section{Microcanonical Ensemble}\label{s7}

As mentioned in the introduction, the third option can be formulated by
using the entropic form  Eq. (\ref{e5}) plus the standard rule  for the
calculation of mean values, Eq. (\ref{e3}) or, alternatively, by using the
original entropic form  Eq. (\ref{e1}), but with a mean value definition:

\begin{equation}\label{e3_NL} 
\langle O\rangle_q = \sum_{i} O_i \frac {p_i^q}{\sum_{j} p_j^q}. 
\end{equation}

These two points of view are completely equivalent. The first option, as
explained in the previous section, uses also the original entropic form but
with standard mean values. We will consider in this section Tsallis
original entropic form $S(q)$ in the  context of the microcanonical
formalism.  The aim is to be able to derive the internal energy without any
{\sl a priori}  assumption about the definition of averages. In the
microcanonical ensemble  we consider the maximization problem for the
original entropic form $S_q$  given by Eq.(\ref{e1}) with the constrain of
given energy $E$.  The solution is the equiprobability, 

\begin{equation}
p_i=\left\{\begin{array}{lr}\Omega(E)^{-1}, &  \varepsilon_i = E \\
0, & {\rm otherwise} \end{array}\right.
\end{equation}

Where $\Omega(E)$ is the number of configurations with energy $E$. The
entropy as a function of the energy is:

\begin{equation}\label{e25}
S_q(E)= \frac {\Omega(E)^{1-q}-1}{1-q} 
\end{equation}

\begin{figure}[!ht]  
\centerline {\epsfxsize=9.5cm \epsfbox{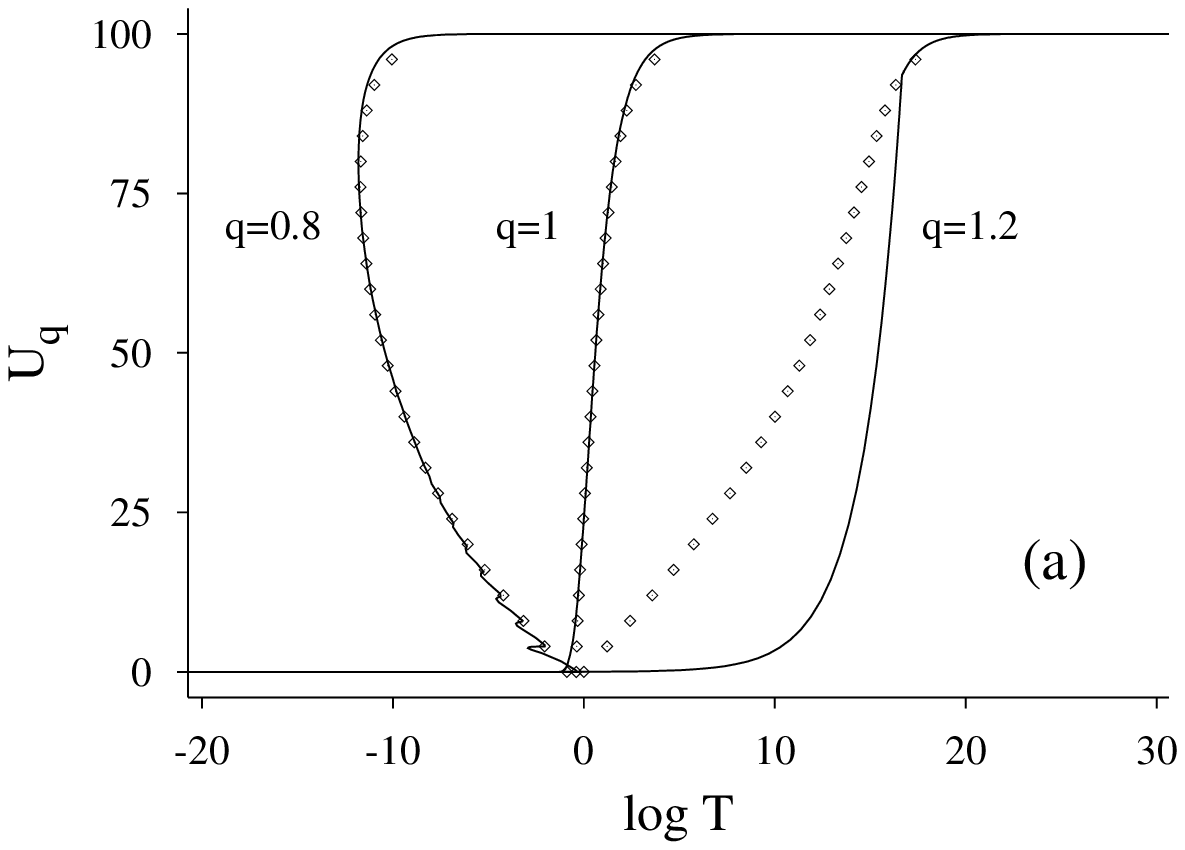}}
\centerline {\epsfxsize=9.5cm \epsfbox{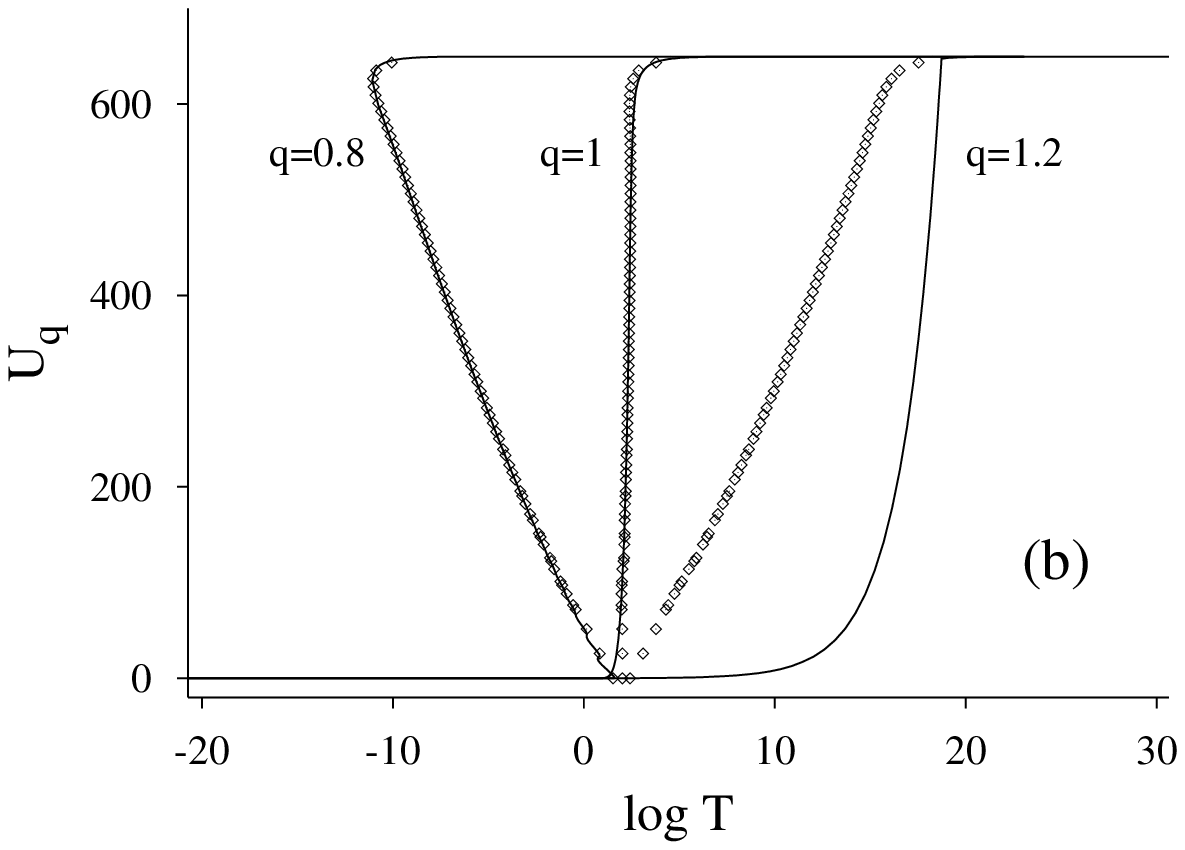}}
\caption{Plot of the internal energy as a function of the temperature for
$N=100$. By points we show the results of using the microcanonical ensemble
described in the text, and by lines the results obtained from the 
canonical ensemble (see Fig. \ref{f7}) by using the non standard  mean
values of the third option. (a) SRIM, (b) LRIM.
\label{f16}}
\end{figure}

The temperature is defined by the thermodynamic relation Eq. (\ref{e8}),
$\frac 1 T= \frac {\partial S_q}{\partial E}$ or

\begin{equation}
\frac 1 T = \Omega(E)^{-q}\frac{\partial \Omega}{\partial E}
\end{equation}

Inverting this relation, we obtain the energy as a function of the 
temperature, $E(T)$. In general, this relation needs to be inverted
numerically. In terms of the scaling function $\phi(x)$ defined in
(\ref{e23}) we have:

\begin{equation}\label{e27}
T = \frac {\tilde N \e^{(q-1)N\phi(E/N\tilde N)}} {\phi'(E/N\tilde N)} 
\end{equation} 

\begin{figure}[!ht]
\centerline {\epsfxsize=9.5cm \epsfbox{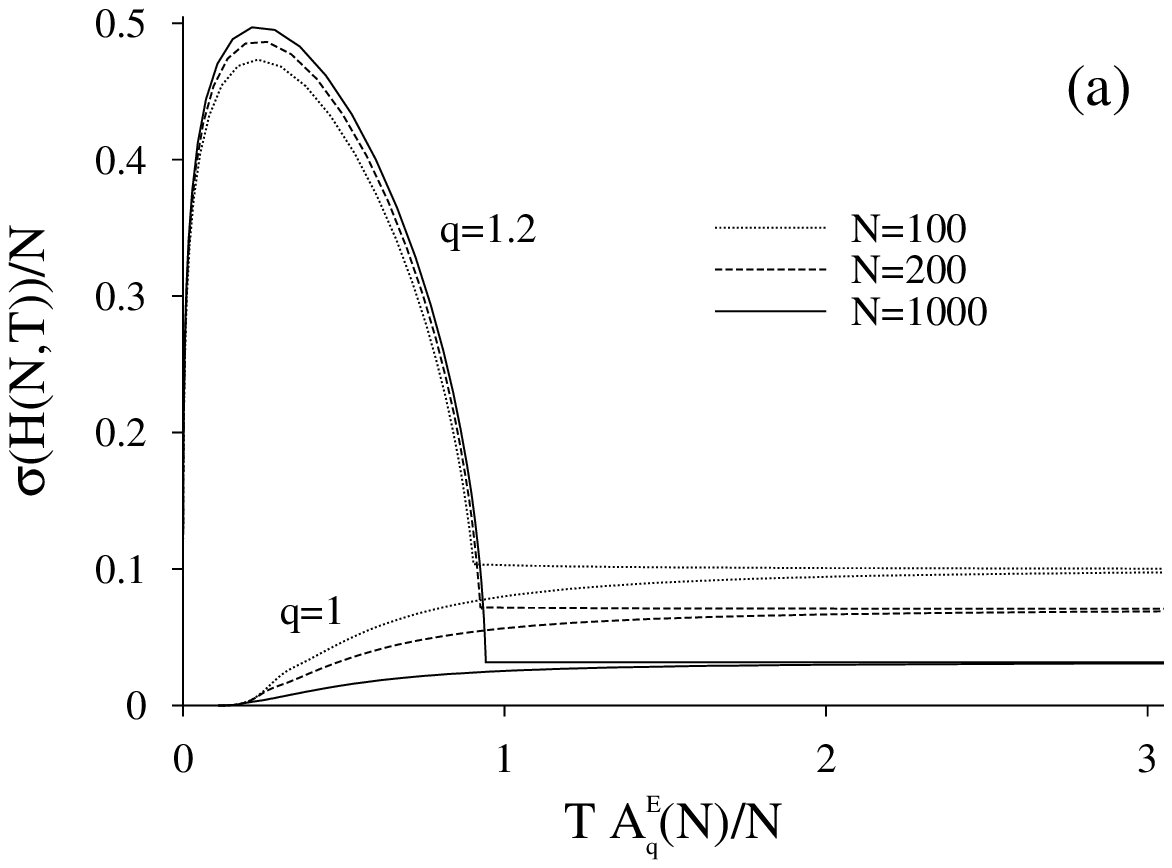}}
\centerline {\epsfxsize=9.5cm \epsfbox{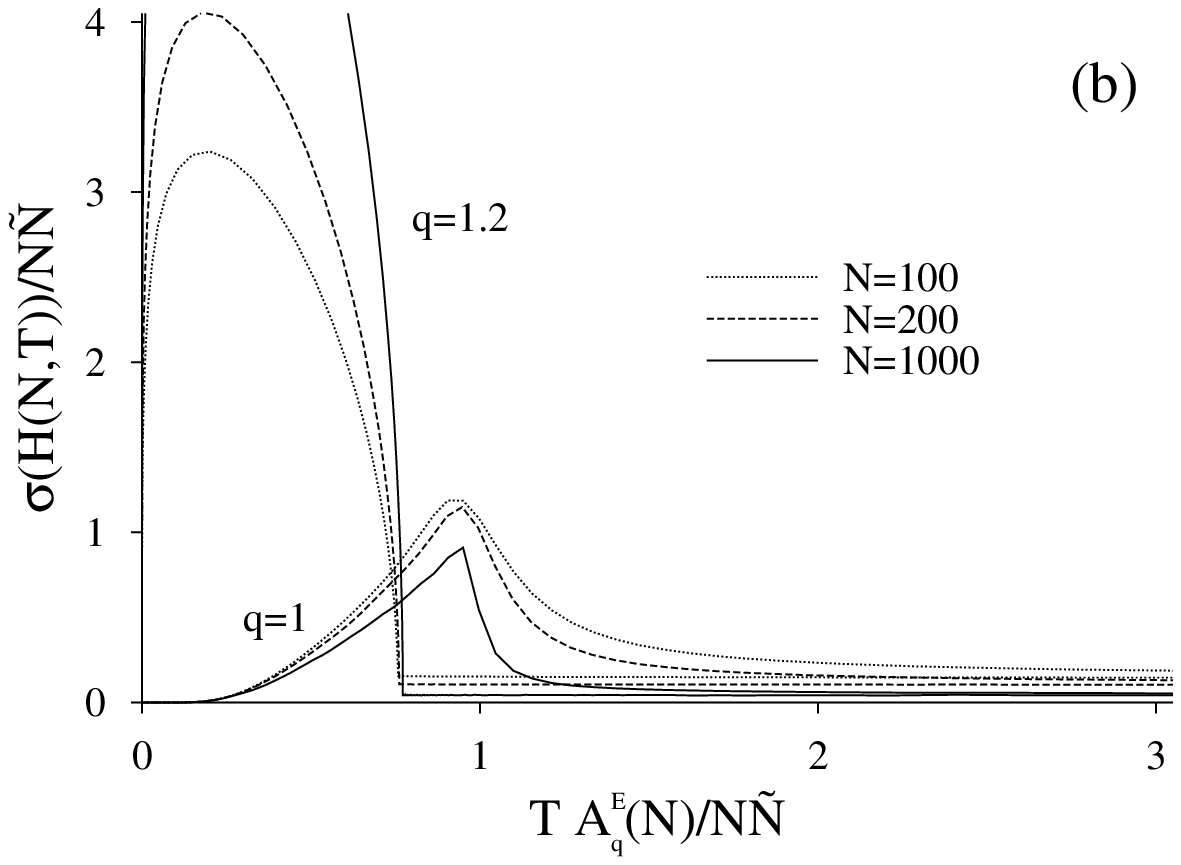}}
\caption{Plot of the energy fluctuations in the canonical ensemble 
$\sigma({\cal H)}/N\tilde N=\sqrt{\langle {\cal H}^2\rangle -\langle {\cal
H}\rangle^2}/N\tilde N$ as a function of the scaled temperature for the
indicated values of the sizes $N$ and the $q$ parameters. (a) SRIM, (b)
LRIM.
\label{f17}}
\end{figure}

where $\phi(x)$ is known exactly for the SRIM and can be evaluated 
numerically using the HOW method for the LRIM (from the plot in the insert
of Fig. \ref{f5}b). Results are shown in Fig \ref{f16} where we plot the 
internal energy coming from the application of the microcanonical ensemble
to both the SRIM and the LRIM. In the same figure we have also plotted the
energy coming from the canonical ensemble using non standard mean values
(third option). We can see that both approaches coincide for $q \le 1$, and
differ for $q > 1$ in some temperature range. The ultimate reason for not
having equivalence between the two ensembles is that fluctuations of the
energy in the canonical ensemble can not be neglected. We have checked that
this is indeed the case by computing the energy fluctuations $\sigma({\cal
H)}=\sqrt{\langle {\cal H}^2\rangle -\langle {\cal H}\rangle^2}$ as a
function of the system size. In the Fig. (\ref{f17}) we see that
fluctuations, normalized by the scale of energy, $N\tilde N$, do not decay
to zero for increasing $N$ in the range of temperatures for which the
microcanonical and canonical ensemble do not agree.  For $q \le 1$
fluctuations do decay to zero with the system size in all the temperature
range.

If we compare the microcanonical and the canonical ensemble using the
standard mean values (the first option studied in the last section) we
observe the coincidence of both ensembles for $q \ge 1$, and disagreement
(in a given temperature range) for $q< 1$. This turns out to be also
consistent with non--vanishing energy fluctuations in the appropriate
range.  This is the expected result because of the mapping $q \to 1/q$
implied in going from the third option to the first one.

\section{Conclusions}\label{s8}

In this paper, we have given details of two methods that can be used to
perform numerical simulations for many particle systems that are governed
by generalized statistics, such as the Tsallis one.  The first method
extends the Histogram by Overlapping Windows method, devised originally for
short range Hamiltonians, to systems with very--long range interactions. 
The second method, devised specifically for the Tsallis thermostatistics,
uses a typical Metropolis Monte Carlo updating scheme combined with a
numerical integration.  We have emphasized the need of using the right
temperature definition if averages are to be independent of the zero of
energy.  We have applied our methods to the case of the Ising model with
either short range (SRIM) or long range (LRIM) interactions. The latter
case corresponds to a situation genuinely non--extensive in which the
energy levels scale as $N^\chi$ with $\chi>1$, $N$ being the number of
variables.  We have compared the methods with some exact results available
in the case of one--dimensional short range Ising model of arbitrary size
and the long--range model for small system sizes.

We have shown that the internal energy, entropy, Helmholtz free--energy and
magnetization follow non trivial scaling laws with the temperature $T$ and
the number of variables $N$. We have justified these scaling laws by some
heuristic arguments that, however, fail to reproduce the observed behavior
for $q>1$. These scaling laws for $q\ne 1$ are non--extensive in the sense
that the different thermodynamic potentials have to be scaled with a factor
that depends in a non--trivial, i.e. non linear, way of $N$. The scaling
laws hold for both the LRIM and the SRIM (with different scaling functions
in each case), independently of the fact that the systems are genuinely
non--extensive or extensive. This shows that the non--extensivity arises
mainly because of the application of the Tsallis statistics. 

We have discussed the differences between the use of standard (first
option) and non--standard (third option) mean value definitions for the
Tsallis Thermostatistics formalism. We show that, although the results of
both definitions can be mapped onto each other by using the $q \to 1/q$
transformations, this mapping requires as well a non--trivial change in the
temperature. Finally, we have shown that the use of the microcanonical
ensemble coincides with the results of the canonical ensemble in the  third
option only for $q\le 1$. We interpret this result as the non--vanishing
energy fluctuations that occur in the corresponding case.  

An obvious extension of the results presented here is to consider the Ising
models in spatial dimension greater than one, \cite{com9}. The HOW method
can be extended in any dimension for short range and long range
interactions. For the SRIM in $d=2$ the exact results \cite{bea96} should be
used.

We remark that the present work concerns equilibrium systems and there is
no time dependence in our simulations. However, it has been recently
conjectured \cite{tsa99} that Tsallis statistics appears in some
non--equilibrium systems such as the relaxation of the non--neutral plasma
experiments in \cite{hua94}. To study these non--equilibrium systems within
the Tsallis statistics formalism,  it would be more appropriate to use
Molecular Dynamics (MD) methods in which the evolution equations are solved
as a function of time. We are currently working on a MD simulation valid
for a Lennard--Jones system within the Tsallis statistics. This MD method
uses a Kusnezov, Bulgac and Bauer thermostat \cite{kus90,pla97} where
additionally the actual temperature has to be calculated using a relation
similar to the Eq. (\ref{e20}).

\noindent Acknowledgments

We wish to thank A.R. Plastino for several discussions about the Tsallis
statistics and for useful suggestions. We acknowledge financial support
from DGES, grants PB94-1167 and  PB97-0141-C02-01.


\begin{thebibliography}{10}
\bibitem{com5}
This affirmation is valid for small $\delta E$ values, which are needed in the
  method used in \cite{lim99}.
\bibitem{com7}
To prove this relation it is useful to consider the following identity: $
  \sum_{j} [1-(1-1/q) \beta_1' \varepsilon_j]^{\frac 1 {q-1}}= (\sum_{j}
  p_j^q)/(1-(1-1/q) \beta_1'\sum_{j} p_j \varepsilon_j) $ .
\bibitem{com9}
A study of the 2-d SRIM using the broad histogram Monte Carlo method was
  carried in \cite{lim99}. .
\end{thebibliography}
\end{document}